\documentclass[12pt, reqno]{amsart}
\usepackage{geometry}                
\usepackage{graphicx}
\usepackage{amsmath}
\usepackage{amssymb}
\usepackage{epstopdf}
\title [Quantum Theory and The Symbolic Dynamics of Invariant Sets] {Quantum Theory \\and The Symbolic Dynamics of Invariant Sets: \\Towards a Gravitational Theory of the Quantum}
\author{T.N.Palmer \\
Clarendon Laboratory, University of Oxford\\
\texttt{tim.palmer@physics.ox.ac.uk}}
\makeatletter
\newcommand\be{\@ifstar{\[}{\begin{equation}}}
\newcommand\ee{\@ifstar{\]}{\end{equation}}}
\newcommand\bp{\begin{pmatrix}}
\newcommand\ep{\end{pmatrix}}

\makeatother
\begin{document}
\bibliographystyle{plain}

\begin{abstract}

A realistic measurement-free theory for the quantum physics of multiple qubits is proposed. This theory is based on a symbolic representation of a fractal state-space geometry which is invariant under the action of deterministic and locally causal dynamics. This symbolic representation is constructed from self-similar families of quaternionic operators. Using number-theoretic properties of the cosine function, the statistical properties of the symbolic representation of the invariant set are shown to be consistent with the contextual requirements of the Kochen-Specker theorem, are not constrained by Bell inequalities, and mirror the statistics of entangled qubits. These number-theoretic properties in turn reflect the sparseness of  the invariant set in state space, and relate to the metaphysical notion of counterfactual incompleteness. Using the concept of probability, the complex Hilbert Space can be considered the completion of this symbolic representation into the state space continuum.  As a result, it is proposed that the complex Hilbert Space should merely be considered a computational convenience in the light of the algorithmic intractability of the invariant set geometry, and consequently the superposed state should not be considered a fundamental aspect of physical theory. The physical basis for the proposed theory is relativistic gravity; for example the symbols used to describe the invariant set themselves label gravitationally distinct cosmological space-times. This implies that the very notion of a `quantum theory of gravity' may be profoundly misguided - erroneously putting the quantum cart before the gravitational horse. Here some elements of an alternative `gravitational theory of the quantum' are proposed, based on a deterministic and locally causal theory of gravity which extends general relativity by being geometric in both space-time and state space. 
\end{abstract}

\maketitle

\section{Introduction}

Although quantum mechanics is generally considered a fundamental theory of physics, it is nevertheless based on the algebraic manipulation of symbols which themselves need have no axiomatic definition. For example,  a state such as $| \text{up}\rangle +  |\text{down}\rangle$ is well defined in quantum theory, even though the symbols `up' and `down' only acquire meaning through the experimental context in which they are used. Consistent with this, Schwinger ~\cite{Schwinger} has derived quantum mechanics as an abstract symbolic theory of `atomic measurements'. 

Symbolic representations of `state' are also common in nonlinear dynamical systems theory ~\cite{LindMarcus} \cite{Williams} \cite{GrabenAtm}.  In the field of Symbolic Dynamics, a system's state space is partitioned into finitely many pieces, each labelled by a distinct symbol. A symbolic representation of the system's state can be defined as a sequence of symbols corresponding to the successive elements of the partition visited by the state in its orbit in state space. This symbolic description can be topologically faithful to the underpinning dynamics. 
 
Are these two notions related? Is it possible to build a theory of quantum physics using ideas from the field of Symbolic Dynamics? Superficially, it would appear not - there exists a plethora of quantum no-go theorems (most famously the Bell Theorem) which would seemingly prevent quantum theory emerging from any realistic locally-causal deterministic theory, symbolic or otherwise. 

In this paper it is argued otherwise, and the key reason is the following. One of the strengths of the symbolic approach to nonlinear dynamics is that, where appropriate, it is able to describe evolution on dynamically invariant subsets of state space, even though geometric properties of such sets may be algorithmically intractable, and, for fractal sets,  formally non-computational in terms of the dynamical equations \cite{Blum} \cite{Dube:1993}.  In this paper, we take a dynamical systems perspective on cosmological space-time \cite{Wainwright}. It is proposed that recent advances in gravitation theory and cosmology argue for a theory of gravity which extends general relativity by being geometric in both space-time and state space, supporting the existence of a fractal invariant set $\mathcal{I}_D$ in the state space of  cosmological space times.  A key postulate of such an extended theory is that the state of the universe lies on $\mathcal{I}_D$ - as discussed this postulate obviates all quantum no-go theorems, determinism and local causality notwithstanding. The purpose of this paper is to show how a realistic measurement-free theory for the quantum physics of multiple qubits is emergent from such an extended theory of gravity and from such a postulate. 

Since $\mathcal{I}_D$ is non-computably related to $D$, the approach taken here is not to try to define a set of differential or difference equations $D$ and deduce from that properties of $\mathcal{I}_D$. Rather, consistent with the notion of quantum theory as a symbolic theory, ideas from the field of Symbolic Dynamics are used to construct directly a representation of  $\mathcal{I}_D$ which emulates the quantum physics of multiple qubits. The existence of $D$ can then be inferred from such a $\mathcal{I}_D$. 

If the physics behind $\mathcal{I}_D$ is essentially gravitational in origin, the essential partitioning of state space which underpins the symbolic approach will be based on the concept of gravitational disimilarity. Conversely, if two elements of state space have the same symbolic label, they are to be considered gravitationally indistinguishable. As discussed in Section \ref{Elements},  this notion is defined using the concept of gravitational interaction energy, made dimensionally consistent and numerically appropriate using Planck's constant. This notion of gravitational indistinguishability provides a point of departure from strict general relativity theory. 

Fractal invariant sets have two key properties: self-similarity and sparseness. Both of these are crucial in constructing the required symbolic representation of $\mathcal{I}_D$. In Section \ref{sequential} it is shown how self-similarity provides a simple way to conceptualise one of the paradigmatic experiments in quantum physics: that of sequential selective spin measurement.  In Section \ref{IST}, a mathematical structure for the symbolic representation of $\mathcal{I}_D$ is developed based on a family of self-similar quaternionic operators acting on symbol sequences: such bit-string symbol sequences are referred to here as `lbits'. This structure not only describes the statistics of sequential spin experiments, it describes more generally the statistical properties of multiple qubits in quantum theory. The framework for the development of such a symbolic representation is referred to generically as `Invariant Set Theory (IST)'.

A key result of this paper is discussed in Section \ref{quantum}: that the abstract complex Hilbert Space of quantum theory can be considered the completion of the symbolic representation of the measure-zero $\mathcal{I}_D$, much like the real numbers form the completion of the measure-zero rationals. The sparseness of $\mathcal{I}_D$ implies that IST is a theory where certain key counterfactual space-times are undefined - we call this `counterfactual incompleteness'. In the symbolic representation, this counterfactual incompleteness manifests itself through number-theoretic properties of the cosine function (e.g. with few exceptions, the cosine of a rational angle is almost always irrational). The Hilbert Space completion - achieved using the concept of probability and well defined on $\mathcal{I}_D$ by simple frequentism - fills in the gaps in $\mathcal{I}_D$ and allows the notion of `state' to be associated with such counterfactual space-times. However, just as unfettered use of the reals over the rationals will lead to unacceptable physical consequences, e.g. as revealed by the Banach-Tarski paradox, it is argued that unfettered use of the complex Hilbert Space will also lead to unacceptable conceptual difficulties: the measurement problem, a particle being `here' and `there' at once, non locality and so on. Hence, it is argued, the Hilbert Space completion of $\mathcal{I}_D$ should not be considered part of fundamental physics, but rather should be considered a calculational convenience in the light of the algorithmic intractability, indeed non-computability, of $\mathcal{I}_D$. 

As discussed in Section \ref{GQ}, this work has relevance not only on the foundations of quantum theory, but also to research into unified theories of physics, and in particular quantum theories of gravity. The physical basis for $\mathcal{I}_D$ is relativistic gravity. For  example, the symbols used to describe $\mathcal{I}_D$ themselves label gravitationally distinct space-times. Here some guidance has been provided by ideas (e.g. \cite{Diosi:1989} \cite{Penrose:2004}) that invoke gravity as a mechanism for objective state reduction in quantum theory (though, emphatically, there is no state reduction in IST). Additionally, the existence of a measure-zero invariant set implies some large (i.e. cosmological)-scale forcing and small-scale irreversibility.  (The notion that forced dissipative turbulent fluids support multi-scale fractal invariant sets provides some motivational guidance for this notion.) Here the positivity of the cosmological constant on the one hand, and information loss at Planck-scale space-time singularities \cite{Penrose:2010} on the other, are invoked. Together this implies that quantum physics may be emergent from a symbolic representation of a deterministic causal theory of gravity that extends general relativity by being geometric not only in space-time, but also in state space. This in turn implies that the very notion of a `quantum theory of gravity' may be profoundly misguided - erroneously putting the quantum cart before the gravitational horse. Here some elements of an alternative `gravitational theory of the quantum' are proposed.

Some potential experimental consequences of IST over quantum theory are discussed in Section \ref{conclusions}. These include an ability to characterise quantum entanglement and weak measurement more completely (both relevant in the field of quantum information), and a prediction that there is no such thing as a `graviton'.

The approach to quantum physics taken in this paper is based on an earlier exploratory study by the author \cite{Palmer:2009a}. 

\section{Preliminaries}
\label{preliminaries}

\subsection{A Symbolic Description of Space-Time Based on Gravitational Similarity}
\label{Elements}

If the space-time in which we live were to be described from a dynamical systems perspective, it would be as a trajectory in some large dimensional Euclidean state space. Such a state space would describe all the `degrees of freedom' needed to distinguish a point on this trajectory from other points on the trajectory.  Consider a portion of this trajectory where distance along the trajectory parametrises some cosmological time (e.g. time since the Big Bang). From its starting point to its ending point, many noteworthy space-time events occur: galaxies collide, stars explode, volcanoes erupt, tropical cyclones make landfall. Each of these events provides a partial description of the space-time trajectory segment. 

Associated with the trajectory portion one can consider even more parochial events: in some laboratory on Earth a source is emitting single quantum particles towards a beam splitter. As shown in Fig 1a, these particles pass through the beam splitter and are registered either by detector A or detector B. Over a long enough trajectory portion the detectors will have registered an ensemble of events large enough to perform some statistical analysis. Focus on one small segment of this trajectory, corresponding to a space-time in which a particle is emitted by the source and detected by A. Once again, from a dynamical systems perspective, one might ask whether the dynamical system $D$ from which this fiducial trajectory is generated, also permits neighbouring state-space trajectories. Indeed, since the world in which we live appears to be chaotic, one might ask for a second state-space trajectory which is initially very close to the fiducial trajectory, but differing in the particle emitted by the source and diverging exponentially from the fiducial trajectory such that now the detector B registers a particle. Indeed one might ask whether there exists an ensemble of diverging trajectory segments, all close to each other and to the fiducial trajectory at initial time, which describe accurately the statistics of detection by A and B, as seen on the much longer fiducial trajectory portion. 

\begin{figure}
\includegraphics{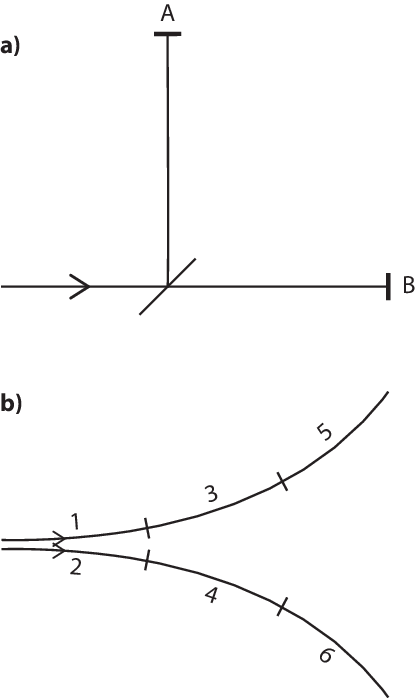}
\caption{a). A source emits single particles which, having passed a beam splitter, are registered either by detector A or detector B. b). Diverging trajectories representing space times associated with a particle emitted by the source and registered by A (top), or with a particle emitted by the source and registered by B (bottom). In trajectory segments 1 and 2, the particles have not yet reached the beam splitter. In trajectory segments 3 and 4 the particles have passed the beam splitter but not yet reached the detectors. By the end of trajectory segments 5 and 6 the particles have initiated a cascade of particle events in detectors A and B respectively. Space-times 1 and 2 are gravitationally deemed gravitationally indistinguishable, as are 3 and 4. Space-times 5 and 6 are dissimilar, as are 1+3+5 and 2+4+6.}
\end{figure}

In this paper, it will be assumed that such a dynamical systems perspective can be applied to the universe in which we live, and that such an ensemble of trajectory segments are consistent with the laws of physics. However, crucially, it will be assumed that such an ensemble is not merely some amorphous collection of trajectories in state space, but respects an underlying measure-zero state-space geometry. This geometry, it will be assumed, is defined by a compact fractal invariant set $\mathcal{I}_D$ associated with the dynamical system $D$ which describes dynamical evolution on these trajectory segments. A guiding example of a fractal invariant set in the analysis below is the multi-scale attractor of chaotic dynamical systems, such as found in studies of fluid turbulence \cite{DoeringGibbon}, though here such sets are considered to relate to the universe as a whole and not to an isolated laboratory system. To make sense of this idea in the present context, we will assume that the universe evolves through an infinity of aeons and hence does not start with the Big Bang (or end in a Big Crunch). The notion of `cyclic' universes is an old one, and even in the case where the universe expands indefinitely into the future, a conformally cyclic cosmology is still possible \cite{Penrose:2010} in which case a compact $\mathcal{I}_D$ is assumed to exist in the conformally rescaled state space. (Here the word `cyclic' is not meant to imply periodic - a fractal $\mathcal{I}_D$ will necessary be aperiodic.) What we humans call `reality' can be associated with a particular trajectory segment on $\mathcal{I}_D$ whose starting point is the Big Bang. Neighbouring trajectories on $\mathcal{I}_D$ represent very similar space-times associated with different aeons either to the remote future of us, or to the remote past. Intelligences which evolve in these remote aeons might in turn refer to these neighbouring trajectories as `reality'. The totality of `reality' is postulated to be precisely $\mathcal{I}_D$.
Thinking of reality on the one hand as defined by an infinite sequence of cosmological aeons, but on the other hand as a single coherent invariant set geometry, embodies the Bohmian notion of explicate and implicate order: two trajectory segments which are distant from the perspective of the explicate order may neighbour each other from the perspective of the implicate order. The physics developed later in this paper draws strongly on this implicate order. 

Leaving aside for now the speculative nature of these remarks, two objections might be raised immediately. The first objection is that at the level of fundamental physics, nature is described by the complex Hilbert Space and hence the sort of `classical' language used above is inappropriate for describing a quantum mechanical system. The second objection is that dynamical systems with fractal invariant sets (such as associated with fluid systems) are forced dissipative systems. Again this seems inappropriate as a description of fundamental physics which is believed to be energy conserving and therefore described by Hamiltonian dynamics. 

These objections can be answered briefly. Firstly, the aim of this paper is to show how the quantum physics of multiple qubits can be developed using the geometric properties of a suitably defined invariant set. Hence the complex Hilbert Space of quantum theory is not considered as fundamental. Secondly, arguments will be made below that the physics needed to generate fractal invariant sets in state space may be gravitational in origin and that there are processes in relativistic gravity which mimic large-scale (i.e. cosmological) forcing and small-(i.e. Planck scale) dissipation. Indeed, the phenomenon of gravity is relevant in defining objectively the notion that  `detector A registers a particle'. Here we will draw on a large body of work which seeks to use gravitation to define objectively `the collapse of the wave function' in quantum theory. However, it is important to add the rider straightaway that the superposed state is not a feature of the theory of quantum physics that will be developed in this paper. 

Fig 1b shows two initially neighbouring trajectory segments. The top trajectory is the fiducial trajectory and describes particle detection by A. The lower trajectory describes particle detection by B.  For reasons to become apparent, the two trajectories have each been split into three segments and the pieces labelled 1-6. Trajectory segments 1 and 2  describe the particles moving some very short distance from the source (from time $t_0$ to time $t_{1,2}$). Trajectory segments 3 and 4 describe the particles passing through the beam splitter (from time $t_{1,2}$ to time $t_{3,4}$) and trajectory segments 5 and 6 describe the passage of the particles from the beam splitter into the detectors and the ultimate registration of the particles by the detectors (from time $t_{3,4}$ to time $t_{5,6}$).

Now two different space-times (e.g. one where a star collapses to form a black hole and one where the star doesn't) can be readily distinguished in terms of their different geometries. Indeed, in classical general relativity one can always distinguish space times from their geometric properties. However, the starting point of our departure from classical physics towards a realistic description of quantum physics is to introduce Planck's constant as a way of determining whether two space-time geometries can in some fundamental sense be considered gravitationally indistinguishable. As a criterion on which this notion of similarity can be based, consider how much energy $E_G$ it would take to move the particles in the first space-time to their position in the second space-time, keeping fixed the gravitational field of the second space time ~\cite{Diosi:1989}  ~\cite{Penrose:2004}). On this basis, two trajectory segments (space-times) of length $\Delta t$ will be said to be gravitationally indistinguishable if 
\be
\label{E_G}
\int_{t}^{t+\Delta t} E_G \; dt < \hbar
\ee
As mentioned, both Di\'{o}si ~\cite{Diosi:1989} and Penrose ~\cite{Penrose:2004} (and references therein) have used $E_G$ (or something like it) to provide an objective criterion for the time $\hbar/E_G$ that a superposed state in standard quantum theory would collapse to a measurement eigenstate under the effects of gravity. 

In the present framework, if (\ref{E_G}) is met, the two trajectory segments will be given the same symbolic labels (e.g. `$a$' or `$b$'). If this condition is not met, the two segments will be given distinct labels (e.g. `$a$' and `$\lnot a$', or `$b$' and `$\lnot b$' etc). For example, it will be assumed that segments 1 and 2 are sufficiently close that criterion (\ref{E_G}) will certainly be met. Similarly, when comparing segments 3 and 4, then although the motion of the original particles through the beam splitter has led to the position and velocity of many individual particles in the beam splitter being different in the two trajectories, leading to some trajectory divergence, it will again be assumed that because gravity is so weak compared with the other `forces' of nature, this divergence is not so great that (\ref{E_G}) is violated. By contrast, segments 5 and 6 correspond to space-times in which a cascade of events has occurred in one or other of the detectors, leading to an enormous number of atoms associated with segment 5 being displaced relative to those in segment 6. For these segments it will be assumed that $E_G$ is violated: order-of-magnitude estimates (\cite{Diosi:1989} \cite{Penrose:2004}) suggest this is reasonable. Whilst trajectory segments 1 and 2 pass the test of gravitational indistinguishability, as do 3 and 4, the combined segments 1+3+5 and 2+4+6 fail the test. 

The notion of $E_G$ can be made rigorous. Gravitational energy momentum is a quasi-local but nevertheless a completely covariant concept and can be defined from tensor fields on the tangent bundle to space-time \cite{Palmer:1978}, \cite{Palmer:1980}. However, in the symbolic approach to $\mathcal{I}_D$ used in this paper, it is not actually necessary to define gravitational energy-momentum explicitly - the symbolism of the approach makes it implicit. 

Of course, space-time contains many particles and hence can potentially be labelled in many different ways. For example, space-time could be given the symbolic label $a_{\mathrm{Alice}}$ if a spin-1/2 particle in Alice's lab is registered by the detector $A_{\mathrm{Alice}}$ in the `up' output of her Stern-Gerlach (SG) apparatus, or the label ${\lnot a}_{\mathrm{Alice}}$ if the particle in Alice's lab is registered by the detector $B_{\mathrm{Alice}}$ in the `down' output of the apparatus. However, space-time could equally be given the label $a_{\mathrm{Bob}}$ if a different particle in Bob's lab is detected by detector $A_{\mathrm{Bob}}$, or the label  ${\lnot a}_{\mathrm{Bob}}$ if the particle is detected by detector $B_{\mathrm{Bob}}$. The labels such as $a_{\mathrm{Alice}}$ and $a_{\mathrm{Bob}}$ associated with neighbouring trajectory segments on $\mathcal{I}_D$ may or may not be correlated. Of course in quantum theory, such correlations can be associated with entangled states. Because of Bell's theorem, such correlations cannot be described by any conventional locally causal hidden-variable theory. A symbolic-dynamic description of the quantum physics of multiple qubits must be able to account for entanglement correlations, and, if $D$ is assumed locally causal, must somehow be able to evade the Bell inequalities. 
 
Returning to this notion of `gravitational indistinguishability', recall that, according to the principle of general covariance in general relativity, there is no natural preferred pointwise identification of two distinct space times. Hence, the notion that (\ref{E_G}) can be used to define gravitationally indistinguishable space-times suggests that the appearance of $\hbar$ in (\ref{E_G}) might signal a breakdown in the principle of general covariance. This in turn is suggestive of some granular structure to space-time on the small scale. In this paper, granularity of space-time is not imposed, but emerges naturally.

Hence if one supposes that if $D$ represents some theory of gravity, it must be an extension of general relativity. Now, as mentioned, certain classes of nonlinear dynamical system, forced dissipative systems, exhibit measure-zero invariant sets in their state space. Fixed points and limit cycles are examples of such invariant sets, but more generically they are fractal.  Here $\mathcal{I}_D$ is considered a fractal invariant set for cosmological space-times: as discussed below, $\mathcal{I}_D$ exemplifies very clearly the Bohmian notion of an `undivided universe' (Bohm and Hiley, ~\cite{BohmHiley}). 

As mentioned above, multi-scale fractal invariant sets are common in the theory of fluid turbulence \cite{DoeringGibbon}. In such systems the forcing is applied at large scales, whilst the dissipation operates on small scales. In dynamical systems theory, such dissipation is manifest in terms of a state-space convergence of trajectories. If such fractal invariant sets have relevance here, the large-scale source of energy must be on the cosmological scale, whilst the small-scale sink of energy must be on Planck scales where the granularity of space-time is apparent. There is indeed a source of large-scale forcing:  the `dark energy' associated with the positive cosmological constant. The notion of a small-scale sink of energy is more controversial. However, there is evidence for it, particularly in the form of black hole information loss. As Penrose ~\cite{Penrose:2010} has emphasised, information loss at a black-hole singularity must be viewed in terms of a convergence of state-space trajectories at the Planck scale. Here, consistent with a proposal by 't Hooft \cite{tHooft} that quantum gravity is dissipative at the Planck scale, a background level of state-space convergence of state-space trajectories is postulated at the Planck scale. The existence of $\mathcal{I}_D$ can be considered as arising from a balance between the large-scale cosmological forcing and the Planck-scale state-space convergence (a rather different perspective on dark energy than provided by conventional physics). 

In the sections below it will be assumed that the universe is evolving on a measure-zero fractal invariant set $\mathcal{I}_D$ in the state space of a (causal deterministic) nonlinear dynamical system $D$. In particular, the Big Bang lies on $\mathcal{I}_D$. One might imagine that the universe has evolved onto $\mathcal{I}_D$ over numerous past cosmic aeons, much as a dynamical system evolves onto its attractor over a very long initial `transient'.  However, the author much prefers the notion that it is simply a primitive law of physics (`The Invariant Set Postulate') that the state of the universe lies on $\mathcal{I}_D$. In keeping with Einstein's great insight, such a putative primitive law of physics is profoundly geometric in origin - though now the geometry is that of state space, in addition to that of space-time. 

Some of the ideas in this Section might be considered rather speculative. However, the author asks the reader not to pass judgement on the proposed theory just yet. In the following Sections, a symbolic representation of $\mathcal{I}_D$ is constructed from which the quantum physics of multiple qubits will emerge. This in turn will lead to some new insights into the nature of the complex Hilbert Space, and the potential for new perspectives on quantum information, weak measurement and quantum gravity. That is to say, the proof of the pudding will be in the eating. 

\subsection{Symbolic Labelling of a Cantor Set and Sequential Selective Measurements in Quantum Physics}
\label{sequential}

Schwinger's symbolic approach \cite{Schwinger} to quantum theory was developed from an examination of sequences of selective measurements. It is therefore appropriate to study such measurements to guide the development of a symbolic representation of $\mathcal{I}_D$. In this section an extremely simplified version of such an experimental set up will be discussed, in order to establish some basic concepts about fractal structure: the more general set up is deferred to later in the paper where the formalism of self-similar quaternion operators is developed. 

A beam of spin-1/2 particles moving in the $\mathbf{\hat{y}}$ direction is fed into an SG apparatus oriented in the $\hat{\mathbf{z}}$ direction. As shown in Fig 2a, the output from which the spin-up particles are sent is absorbed by A, and the output from which the spin-down particles are sent is fed into a second SG apparatus. In this simplified version of the experiment, this second apparatus can be oriented in only one of four ways: in the $\pm \hat{\mathbf{x}}$ directions and in the $\pm \hat{\mathbf{z}}$ directions. The spin-up output is absorbed by B and the spin-down output is fed into a third SG apparatus which again can only be oriented in one of the  $\pm \hat{\mathbf{x}}$ and $\pm \hat{\mathbf{z}}$ directions. Spin-down and spin-up output from the third apparatus is registered by detectors C and D respectively. 
 \begin{figure}
\includegraphics[scale=0.7] {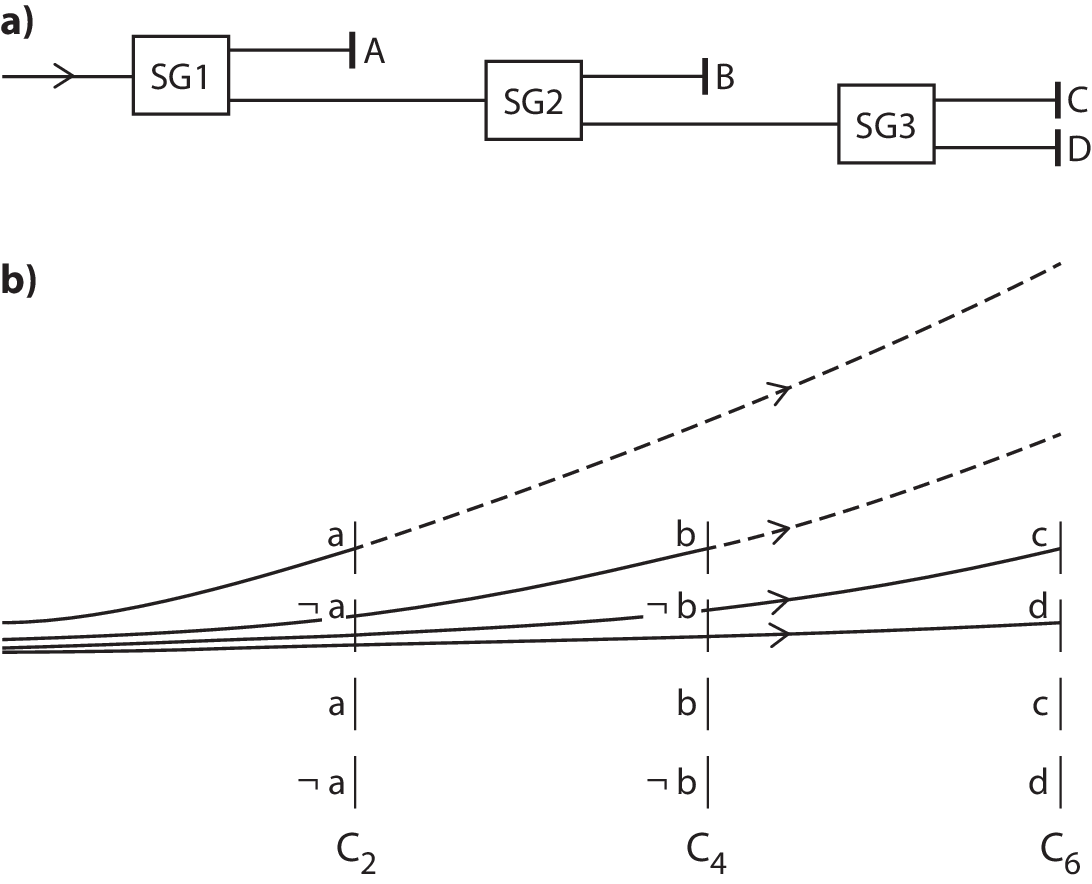}
\caption{a). Three sequential Stern-Gerlach spin measurements. b) State-space trajectories (space-times) associated with particles emitted by the source and registered by A , B, C and D respectively. Shown are single groupings of elements of the iterates $C_2$, $C_4$ and $C_6$ of the Cantor Set $C$ of trajectories comprising $\mathcal{I}_D$. $C_2$, $C_4$ and $C_6$ have been drawn at different times, corresponding to the times $t_1, t_2, t_3$ by which the positive Lyapunov exponents on $\mathcal{I}_D$ have magnified the structure of $C_2$, $C_4$ and $C_6$ to the same state-space scale. The four trajectories can be represented by a symbolic string associated with the elements of the three iterates to which they belong: in this case $.aaa$, $.\lnot a b b$, $.\lnot a \lnot b c$, and $.\lnot a \lnot b d$ respectively. As in standard symbolic dynamic representations, time evolution is effected by shifting the radix point one step to the right, and erasing the symbol to the left of the radix point.}
\end{figure}
As is well known, if the second apparatus is oriented in the $\hat{\mathbf{x}}$ direction and the third in the $\hat{\mathbf{z}}$ direction, then C is found to detect particles, despite the fact that these spin-up particles should, from any realist classical perspective, have all been absorbed by A. It seems that the second SG apparatus has somehow destroyed any previous information about the particles' spin characteristics in the $\hat{\mathbf{z}}$ direction. 
 
Let us approach this problem from the symbolic-dynamic perspective. Consider the state-space trajectories illustrated in Fig 2b. Four neighbouring trajectories are shown starting at $t_0$, and diverging from each other with time. As discussed above, it will be supposed that these four trajectories lie on some fractal set $\mathcal{I}_D$ in state space. The top trajectory describes a space-time where a particle emitted by the source passes through the first SG apparatus and is registered by $A$ at time $t_1$. Although the space-time does not end at $t_1$, we lose interest in it at $t_1$ and so beyond $t_1$ it is shown as dotted. The lower three trajectories describe space-times where a particle emitted by the source passes through the first SG apparatus and is input to the second SG apparatus. Based on the segment length $t_1-t_0$ and the discussion above, the top trajectory is gravitationally dissimilar to any of the lower three (i.e. the condition (\ref{E_G}) fails). The top trajectory can therefore be labelled $a$ and the bottom three $\lnot a$.
 
 Focus now on these three lower trajectories labelled $\lnot a$. The uppermost of these corresponds to a space-time where a particle emitted by the source is registered at $B$. Based on the intermediate segment length $t_2-t_0$, this trajectory can be labelled $b$ and is gravitationally dissimilar to either of the lower two which can therefore, in addition to having the label $\lnot a$, has the label $\lnot b$. As before, we lose interest in the $b$ trajectory after $t_2$. Of the two remaining two trajectories, the top describes a space-time where a particle emitted by the source is registered by $C$ at $t_3$. Over the segment length $t_3-t_0$ this trajectory, in addition to having the labels $\lnot a$ and $\lnot b$, has the label $c$. The bottom trajectory is registered at D and therefore, in addition to having the labels $\lnot a$ and $\lnot b$, has the label $d$.  
 
It is now shown: a) how the symbolic description of these four trajectories can be shown to arise naturally from a symbolic representation of a fractal set, and b) how a random sampling of trajectories from this fractal set is consistent with the statistics of sequential measurements in quantum physics for the limited set of orientations allowed for the SG devices in this section.  

If $\mathcal{I}_D$ is a fractal set of state-space trajectories, each a space-time, then $\mathcal{I}_D$ is locally the Cartesian product of $\mathbb{R}^1$ with some sort of Cantor Set $C$. A Cantor set is the limit of a self-similar iterative process. For the simplest Cantor Ternary Set
\be
\label{Cantor}
CT=\bigcap_{k \in \mathbb{N}} CT_k 
\ee
 where $CT_0$ is the unit interval, $CT_1$ is obtained by removing the `middle third' from the unit interval, $CT_2$ is defined by removing the middle third from each of the two pieces of $CT_1$, etc. That is, the $k+1$th iterate $CT_{k+1}$ comprises two copies of $CT_k$, each copy being reduced by a factor of $1/3$ and separated by a gap of width $1/3$ relative to the size of $CT_k$. The fractal dimension of the limit set $CT$ is $\log{2}/\log{3} < 1 $. 

$CT$ is but one type of Cantor Set. A family of alternate Cantor Sets $C^{(M)}$ of more relevance here is defined for any $M \in \mathbb{N}$, such that $C^{(M)}_{k+1}$ comprises $2^M$ uniformly spaced copies of $C^{(M)}_k$, each reduced in size by a factor of $1/2^{M+1}$. Each copy is separated from its neighbour by a gap of size $1/2^{M+1}$ relative to the size of $C^{(M)}_k$.  The fractal dimension of $C^{(M)}$ is $M/(M+1) \rightarrow 1$ as $M\rightarrow \infty$. For the rest of this section, the superscript on $C^{(M)}$ will be dropped. 

To illustrate how to link the sequential selective SG measurements to a symbolic labelling of a Cantor Set, we will take $M=2$ . This provides only a `toy model' of the full invariant set, but it nevertheless has enough structure to be able to describe some of the basic concepts of IST (though $M=2$ is far too small to be able to describe either the apparent stochasticity of quantum physical measurements and the full range of measurement orientations). With $M=2$, the first iterate $C_1$ of $C$ comprises $4$ copies of the unit interval $[0,1]$, each copy reduced by a factor $1/8$. The word `element' will be used to describe one of these copies. Hence, for the second iterate $C_2$, each of the 4 elements of $C_1$ itself comprises 4 further elements, reduced again by a factor $1/8$ relative to $C_1$. So $C_2$ can be described as comprising 4 groupings, each of 4 elements. 

The fractal structure of $\mathcal{I}_D$ is inherently linked with the chaotic dynamics of $D$. In particular, positive Lyapunov exponents associated with $D$ describe the exponential rate of divergence of neighbouring trajectories on $\mathcal{I}_D$.  After an e-folding time $T_L$ given by these positive Lyapunov exponents, the divergence of trajectories on $\mathcal{I}_D$ will lead the finer-scale elements associated with $C_2$ will be magnified to the coarse-scale of the $C_1$ elements. After a further e-folding time, the even finer-scale elements of $C_3$ elements will be amplified onto the coarse-scale of $C_1$ - and so on. That is to say, the self-similar structure of $C$ can be viewed in two different ways: firstly in a dynamically passive way where `zooming' into some static version of $C$ reveals its finer-scale iterates one by one, and secondly in a dynamically active way where the positive Lyapunov vectors magnify the higher iterates of $C$ one at a time onto some fixed coarse scale. 

A scheme is now developed whereby the iterates of $C$ are labelled consistent with the statistics of the sequential measurements above. Consider the iterate $C_1$. By construction it will be assumed that for short times to the future of $t_0$, all state-space trajectories associated with any one of the 4 elements of $C_1$ will be gravitationally indistinguishable:  more specifically that 
\be
\label{E_G2}
\int_{t_0}^{t_0+T_L} E_G \; dt < \hbar 
\ee
for all trajectories of length $T_L$, through any of the four individual elements of $C_1$. Hence label each element of $C_1$ with the label of these state-space trajectories. On the other hand, it is not assumed that trajectories through different elements of $C_1$ are gravitationally indistinguishable. That is, different elements of $C_1$ can have different labels (such as `$a$' and `$\lnot a$'). 

Now consider the second iterate $C_2$. Each element of $C_1$ itself comprises four elements in $C_2$. If the trajectories through individual elements of $C_1$ satisfy (\ref{E_G2}), then it is assumed that the trajectories through the individual finer-scale elements of $C_2$ will satisfy
\be
\label{E_G3}
\int_{t_0}^{t_0+2T_L} E_G \; dt < \hbar \nonumber
\ee
Hence label each element of $C_2$ by the label of the associated space-time trajectories over $2T_L$. Again, different elements of $C_2$ can have different labels. 

In general, assume the trajectories through each of the individual elements of $C_k$ satisfy
\be
\label{E_Gi}
\int_{t_0}^{t_0+IT_L} E_G \; dt < \hbar \nonumber
\ee
Hence label each element of $C_k$ by the label of the associated space-time trajectories through that element, over the time $kT_L$. A particular point of $C$ is then defined by a sequence of labels, one for the element of each iterate to which the point belongs. 

In Section \ref{quaternion}, a general mathematical ansatz is developed for labelling the elements of the iterates of $C$. It is based on families of self-similar square-root-of-minus-one operators. If this ansatz is applied to the 16 elements of $C_2$, arranged as 4 groupings of 4 elements, it gives 
 \be
 \label{SGlabel2}
  a \; \lnot a \;  a \; \lnot a \;\;\;\;\;\;\;\; \lnot a \; \lnot a \; \lnot a \; \lnot a \;\;\;\;\;\;\;\;
\lnot a \; a \; \lnot a \;  a \;\;\;\;\;\;\;\;  a \;  a \;  a \;  a
  \ee

This labelling ansatz can be related to the problem of sequential spin measurements. If one focusses on the fourth of the groupings of labels in (\ref{SGlabel2}), it can be seen that all elements are labelled `$a$'. If any one of these elements is selected from this fourth grouping, it has the label given to a trajectory associated with a space-time where a particle is absorbed by A. Instead, if one focusses on the second of the groupings of labels in (\ref{SGlabel2}), and if any one of the elements is selected, it has the label given to a trajectory associated with a space-time where a particle is output to the second SG device. Finally, if one focusses on either the first or third grouping, there is a probability of 1/2 of selecting an element with either an `$a$' label or a `$\lnot a$' label. The four trajectories in Fig 2b at $t_1=2T_L$ can arise from sampling either the first or third groupings in $C_2$. The probability of sampling one `$a$' and three `$\lnot a$'s (i.e. the labels of the four trajectories shown in Fig 2b) from these groupings is equal to 1/4. 

By self similarity, each element of $C_2$ is associated with 16 elements of $C_4$ (4 groupings of 4 elements). As described above, the labelling of the elements of $C_4$ is associated with the labels of space-time trajectories over the time interval $t_2=4T_L$.  Since we are not interested in the $a$ trajectories after $t_1$ (where the particle has been absorbed by A), we focus on those elements of $C_4$ which are associated with an element of $C_2$ that has been labelled $\lnot a$. In the ansatz of the next Section, the labelling of these particular elements of $C_4$ is also given by (\ref{SGlabel2}), i.e. 
\be
 \label{SGlabel4b}
  b \; \lnot b \; b \; \lnot b \;\;\;\;\;\;\;\; \lnot b \; \lnot b \; \lnot b \; \lnot b \;\;\;\;\;\;\;\;
\lnot b \; b \; \lnot b \;  b \;\;\;\;\;\;\;\;  b \;  b \;  b \;  b
  \ee
except that the `$b$' label replaces the `$a$' label. 

For the four individual groupings in (\ref{SGlabel4b}), the probability of selecting `$b$' is equal 1/2, 0, 1/2 and 1 respectively. These four values correspond to the probabilities that $B$ registers a particle when the second SG apparatus is oriented in the $\mathbf{\hat{x}}$, $\mathbf{\hat{z}}$, $-\mathbf{\hat{x}}$, and $-\mathbf{\hat{z}}$ directions respectively. Let us suppose that indeed the second apparatus is oriented in the $\mathbf{\hat{x}}$ direction. Then the probability of selecting one `$b$' trajectory and two `$\lnot b$' trajectories from the first grouping is 3/8.

By self similarity, each element in $C_4$ is associated with 16 elements in $C_6$ (4 groupings of 4 elements).  As described above, the labelling of the elements of $C_6$ is associated with the labels of space-time trajectories over the time interval $t_2=6T_L$.  Since we are neither interested in the $a$ trajectories after $t_1$, nor the $b$ trajectories after $t_2$, we focus on those elements of $C_6$ which are associated with an element of $C_4$ that has been labelled `$\lnot b$'.  The labelling of these particular elements of $C_6$ is identical to $C_4$ i.e. 
\be
 \label{SGlabel6c} 
  c \; d \; c \; d \;\;\;\;\;\;\;\; d \; d \; d \; d \;\;\;\;\;\;\;\;
d \; c \; d \; c \;\;\;\;\;\;\;\;  c \;  c \;  c \;  c
  \ee
except that the `$c$' label replaces the `$b$' label, and the `$d$' label replaces the `$\lnot b$' label. The probabilities associated with the four groupings correspond to situations when the third SG apparatus is oriented in the $\mathbf{\hat{z}}$, $\mathbf{\hat{x}}$, $-\mathbf{\hat{z}}$, and $-\mathbf{\hat{x}}$ directions respectively. For example, if the third SG apparatus is oriented in the $\mathbf{\hat{z}}$ direction, the probability that C registers a (spin-up) particle is equal to 1/2 based on the first grouping in (\ref{SGlabel6c}). This was the situation difficult to understand with a simple classical realist model. It is easy to emulate in this fractal framework. 

What does this labelling imply about the underlying dynamics $D$? As discussed, the labellings have been defined on the basis of gravitational indistinguishability (or, conversely, dissimilarity). In some sense, different labels can be thought of as defining different gravitational basins of attraction on $\mathcal{I}_D$. The fact that neighbouring elements of $C$ can have distinct labels is precisely what one would expect in the theory of riddled basins of attraction in chaotic dynamical systems theory \cite{Alexander:1992}. A system has a riddled basin of attraction if the neighbourhood of a point in a particular basin of attraction contains points which are not in that basin of attraction. 

As discussed, the points of $C$ can be represented by sequences of symbols associated with the symbolic labelling of the iterates to which the point belongs. The four trajectories illustrated in Fig 2 can therefore be represented by the symbolic sequences $.aaa\ldots$, $.\lnot a b b \ldots$, $.\lnot a \lnot b c \ldots$, and $.\lnot a \lnot b d\ldots$, where the $I$th label is taken from the labelling of the element of $C_{2I}$ to which the trajectory belongs. As in standard symbolic dynamic representations, time evolution (e.g. $t_1 \rightarrow t_2 \rightarrow t_3$) is effected by shifting the radix point one step to the right, and erasing the symbol to the left of the radix point. 

With this toy model in mind, a general formulation for the symbolic labelling of $\mathcal{I}_D$ is now developed. 

\section{Invariant Set Theory}
\label{IST}

In an earlier exploratory paper ~\cite{Palmer:2009a}, the author introduced the `Invariant Set Postulate' to describe the notion that states of physical reality lie on some measure-zero subset of state space. Here, the ideas introduced in this earlier paper are developed and given some quantitative substance. As such, the basic structure outlined below will now be described as `Invariant Set Theory' (IST). 

As discussed in the introduction, if one were to take the standard route in dynamical systems theory, one would first define $D$ as a self-contained mathematical system (e.g. based on differential or finite-difference equations) and infer from it the structure of $\mathcal{I}_D$. However, such an approach founders at the first step: the geometric properties of fractionally dimensioned $\mathcal{I}_D$ are, in general, not computably related to $D$ (see \cite{Blum}). For example, there is no finite algorithm based on $D$ for determining whether a given point in state space lies on a fractal invariant set $\mathcal{I}_D$, nor, for example, whether a given line intersects $\mathcal{I}_D$ \cite{Dube:1993}.  In fact, given the unimaginably large dimension of the Euclidean state space needed to embed $\mathcal{I}_D$, the determination of any geometric aspect of $\mathcal{I}_D$ from $D$ will be algorithmically intractable. Of course this is entirely consistent with the notion that experimenters such as Alice and Bob, who are surely part of the universe and hence subject to the same laws of physics as the particles they study, can for all practical purposes by considered `free agents', despite the presumed determinism of $D$ (see \cite{Lloyd:2012} and further discussion in Section \ref{EI} below). 

As a result, the development here is motivated by a different logic, consistent with the notion that it is the geometry of $\mathcal{I}_D$ that is fundamental and the equations $D$ secondary. That is to say we ask: Given that quantum mechanics is itself a symbolic theory, can a symbolic representation of an invariant-set geometry $\mathcal{I}_D$ be constructed without explicit reference to $D$, such that the quantum physics of multiple qubits is emulated by the statistical properties of this symbolic representation of $\mathcal{I}_D$? This approach is not so much different from those where the invariant set of some dynamical system is inferred from empirical time series. Symbolic dynamics is frequently used for such `attractor reconstructions'. See for example ~\cite{Gilmore}.

\subsection{Symbolic Labelling of $\mathcal{I}_D$ Based on Self-Similar Families of Quaternions}
\label{quaternion}

With the toy model of the last section in mind, consider a multi-dimensional generalisation of the Cantor Set of the previous section. In the following discussion, the parameter $N$ is itself a power of 2.

\subsubsection{Symbol Sequences and Co-Sequences}

Our starting point is the sequence  
\be
 ( a |=(a \; a \; a \; \ldots \;  a) \nonumber \\
 \ee
of `$a$' symbols, of length $2^N$. Such a sequence will also be represented in co-sequence form, i.e. 
  \be
  \label{aaa}
 |a ) = 
 \left ( \begin{array}{c}
  a \\ a \\ a \\ \vdots \\ a 
\end{array} \right ) \nonumber
\ee
also of length $2^N$. From this we can define the negation operator
\begin{align}
 -( a |&=(\lnot a \; \lnot a \; \lnot a \; \ldots \; \lnot a) \nonumber 
 \end{align} 
 and
 \be
- |a ) = 
 \left ( \begin{array}{c}
  \lnot a \\ \lnot a \\ \lnot a \\ \vdots \\ \lnot a 
\end{array} \right ) \nonumber
\ee
The labelling of the iterates of $\mathcal{I}_D$ is obtained by operating on $|a)$ with operators $\mathcal{\bar{U}}$ (of which the negation operator `-' is an example). In the development below, it will be convenient to represent the operators $\mathcal{\bar{U}}$ by $2^N \times 2^N$ matrices where in each row and column is full of `0's except for one entry which is either equal to `1' or to `$\lnot$', the identity and negation operators respectively. That is, 
 \begin{align}
&1 (a) = a; \;\;\; 1(\lnot a) =\lnot a \nonumber \\
&\lnot(a) = \lnot a; \;\; \lnot(\lnot a) = a \nonumber
\end{align}
 The `null' operator `0' can be trivially defined with the properties that   
\begin{align}
(1+0) (a) = (0+1) (a) = a \;\;\; &(\lnot+0)(a) = (0+\lnot)(a)= \lnot a \nonumber \\
(1+0) (\lnot a) = (0+1) (\lnot a) = \lnot a \;\;\; &(\lnot+0)(\lnot a) = (0+\lnot)(\lnot a)= a \nonumber
\end{align}
and will henceforth be ignored by blanking out all occurrences of `0' elements in matrix representations of operators, e.g. 
\be
\label{example}
 \mathcal{U}=
 \bp 0 & 1& 0& 0\\ \lnot &0&0&0\\0&0&0&\lnot\\0&0&1&0 \ep
 \equiv
 \bp \; & 1& \;& \;\\ \lnot&\;&\; &\;\\\;&\;&\;&\lnot\\\;&\;&1&\; \ep 
 \ee
That is to say, in the operators constructed below, there is exactly one non-zero operation. An important piece of notation is now defined. Given an $nN \times nN$ matrix operator $\mathcal{U}$, then $\mathcal{\bar{U}}$ is defined as the $2^N \times 2^N$ matrix operator
 \be
 \label{bar}
 \mathcal{\bar{U}}=
\bp \mathcal{U} & \;&\;&\; \\ \;& \mathcal{U}& \;&\; \\ \;& \;& \ddots \; \\ \;& \;& \; &\; & \mathcal{U} \ep.
 \ee 
containing $2^N/nN$ copies of $\mathcal{U}$. Hence, with the specific example (\ref{example})
\be
\bar{\mathcal{U}} |a)=
\left ( \begin{array}{c}
  a\\ \lnot a \\ \lnot a \\ a \\ \vdots  
\end{array} \right ) \nonumber
\ee
where the 16-element co-sequence $\mathcal{\bar{U}} |a)$ repeats every 4 elements. 

\subsubsection{Square Roots of Minus One} 

From the $2^0 \times 2^0$ operators $1$ and $\lnot$, define the $2^1 \times 2^1$ permutation/negation operators 
 \be
 \label{i}
1= \bp 1 &\;\\\;& 1 \ep;
\;\; -1=\bp \lnot & \; \\ \;& \lnot \ep; \nonumber \\
 \;\;i=  \bp \; & 1 \\ \lnot & \; \ep;
\;\; -i= \bp \;& \lnot \\ 1  &\; \ep.
 \ee
To avoid a proliferation of symbols, the symbol `1' is used to denote the identity matrix irrespective of matrix size. The implied size of `1' should be obvious from the context. Hence, 
 \be
 i \circ i = i^2  = -1 \nonumber
 \ee
 so that $i$ can be treated as a square root of $-1$. (It can be noted in passing that the binary labelling of the 4 groupings of symbol elements discussed in Section \ref{sequential} is given by the co-sequences
  \be
  \label{fig2b}
  \bar{i}|a)=
  \left ( \begin{array}{c}
  a\\ \lnot a \\ a \\  \lnot a 
\end{array} \right ); \;
  \bar{i}^2 |a)=
   \left ( \begin{array}{c}
  \lnot a\\ \lnot a \\ \lnot a \\ \lnot a  
\end{array} \right ); \;
 \bar{i}^3 |a)=
 \left ( \begin{array}{c}
  \lnot a\\ a \\ \lnot a \\ a 
\end{array} \right ); \;
 \bar{i}^4 |a) =
 \left ( \begin{array}{c}
  a\\ a \\ a \\  a 
\end{array} \right )
  \ee
in the simplest possible `toy' universe with $N=2^1$.

From these $2^1 \times 2^1$ operators, define the $2^2 \times 2^2$ permutation/negation operators
\be
\label{three}
\mathbf{E}_0=\bp \;&i\\i&\;\ep;
\;\;\mathbf{E}_1=\bp i&\;\\\; &-i\ep;
\;\;\mathbf{E}_2=\bp \;& 1\\ -1 &\; \ep
\ee
Importantly, the $\mathbf{E}_j$ satisfy the familiar rules for quaternionic multiplication; not only does $\mathbf{E}^{2}_0=\mathbf{E}^{2}_1=\mathbf{E}^{2}_2=-1$, but also
\be
\label{quaternions}
\mathbf{E_0} \circ \mathbf{E_1}= \mathbf{E}_2
\ee
Using the notion of self-similarity, the operators $\{\mathbf{E}_0, \mathbf{E}_1\}$ can in turn be used as block matrix elements to generate the four $2^3 \times 2^3$ square-root-of-minus-one operators. 
\begin{align}
\label{seven}
\mathbf{E}_{00}=\bp \; &\mathbf{E}_0 \\ \mathbf{E}_0& \; \ep;
\;\;\mathbf{E}_{01}&=\bp \; & \mathbf{E}_1 \\ \mathbf{E}_1& \; \ep \\
\mathbf{E}_{10}=\bp \mathbf{E}_0 & \;\\ \; & -\mathbf{E}_0 \ep;
\;\;\mathbf{E}_{11}&=\bp \mathbf{E}_1 & \;\\ \; & -\mathbf{E}_1 \ep
\end{align}
which satisfy the following quaternionic relationships:
\begin{align}
\label{morequats}
\mathbf{E}_{00} \circ \mathbf{E}_{10} = 
\mathbf{E}_{01} \circ \mathbf{E}_{11} = \bp \;& 1\\ -1 &\; \ep \\ \nonumber
\end{align}
The permutation/negation operators in (\ref{seven}) can be ordered in the set
\be
\{\mathbf{E}_{00}, \mathbf{E}_{01}, \mathbf{E}_{10}, \mathbf{E}_{11} \}
\ee
and can be used to generate the eight $2^4 \times 2^4$ square-root-of-minus-one operators
\begin{align}
\label{fifteen}
\mathbf{E}_{000}=\bp \; &\mathbf{E}_{00} \\ \mathbf{E}_{00}& \; \ep;  
\;\;\mathbf{E}_{001}=&\bp \;& \mathbf{E}_{01} \\  \mathbf{E}_{01} & \; \ep;
\dots
\;\;\mathbf{E}_{011}=\bp \; & \mathbf{E}_{11} \\ \mathbf{E}_{11} & \; \ep\\ \nonumber
\mathbf{E}_{100}= \bp \mathbf{E}_{00} & \;\\ \; & -\mathbf{E}_{00} \ep;
\;\;\mathbf{E}_{101}=&\bp \mathbf{E}_{01} & \; \\ \;& -\mathbf{E}_{01} \ep;
\dots
\mathbf{E}_{111}=\bp  \mathbf{E}_{11} &\; \\ \;& -\mathbf{E}_{11} \ep 
\end{align}
which satisfy quaternionic relationships
\be
\mathbf{E}_{000} \circ \mathbf{E}_{100} = \mathbf{E}_{001} \circ \mathbf{E}_{101}= \ldots = \mathbf{E}_{011} \circ \mathbf{E}_{111}=\bp \;& 1\\ -1 &\; \ep
\ee
In turn, these operators form an ordered set
\be
\label{seq}
\{\mathbf{E}_{000}, \mathbf{E}_{001}, \mathbf{E}_{010}, \mathbf{E}_{011}, \mathbf{E}_{100}, \mathbf{E}_{101}, \mathbf{E}_{110}, \mathbf{E}_{111}\}
\ee
and can be used to generate 16 $2^5 \times 2^5$ square-root-of-minus-one operators, and so on to quaternions associated with a square matrix whose order is any power of 2. 

Let $\beta$ denote a string of `$0$'s and `$1$'s, then it is trivially shown that
\begin{align}
\mathbf{E}_{0\beta}&=\bp \mathbf{E}_\beta  \; \\ \; & \mathbf{E}_\beta & \ep \sigma_x \\
\mathbf{E}_{1\beta}&=\bp \mathbf{E}_\beta  \; \\ \; & \mathbf{E}_\beta & \ep \sigma_z
\end{align}
where 
 \be
 \label{pauli}
\sigma_x= \bp \: &1\\1& \; \ep;
 \;\;\sigma_y=  \bp \; & -\mathbf{E}_\beta \\ \mathbf{E}_\beta & \; \ep;
\;\; \sigma_z = \bp 1& \; \\ \;  &-1 \ep.
 \ee
can be defined as Pauli permutation/negation operators, relating in an obvious way to conventional complex Pauli matrices. 

If we insert a radix point after the first digit in each of the subscript sequences in, for example, (\ref{seq}), then the ordered set of independent quaternion operators can be written as 
\be
\label{sequence}
\{\mathbf{E}_{\beta}\}
\ee
where now $0 \le \beta < 2$ is a dyadic rational (instead of a bit string). The negation of the operators (\ref{fifteen}) are contained in the set
\be
\{-\mathbf{E}_{0.00}, -\mathbf{E}_{0.01}, -\mathbf{E}_{0.10}, -\mathbf{E}_{0.11}, -\mathbf{E}_{1.00}, -\mathbf{E}_{1.01}, -\mathbf{E}_{1.10}, -\mathbf{E}_{1.11}\}
\ee
which can be appended to (\ref{sequence}) by putting
\be
\{-\mathbf{E}_{0.00}, -\mathbf{E}_{0.01}, -\mathbf{E}_{0.10}, \ldots -\mathbf{E}_{1.11}\} \rightarrow \{\mathbf{E}_{10.00}, \mathbf{E}_{10.01}, \mathbf{E}_{10.10}, \ldots \mathbf{E}_{11.11}\} 
\ee
In the following, the symbolic labelling of $\mathcal{I}_D$ will be based on the square root of minus one operators $\{\mathbf{E}_{\beta}\}$ where $0 \le \beta <4$ and where the corresponding matrices have have size $N \times N$ (where $N$ is a power of 2). Hence $\beta$ is a dyadic rational describable by $\log_2 N +1$ bits (the `$+1$' denoting the sign bit). For $N=4$, Fig 3 shows the sequence $\{\mathbf{E}_\beta\}$ arranged on a circle. Note that operators at diametrically opposite points are the negation of one another. Let $\mathbb{Q}_2$ denote the set of dyadic rationals and $\mathbb{Q}_2 (M)$ the set of dyadic rationals described by $M$ bits. Then $\beta \in \mathbb{Q}_2(\log_2 N +1)$. Below we will consider the limit $N \rightarrow \infty$. 
\begin{figure}
\includegraphics[scale=0.3] {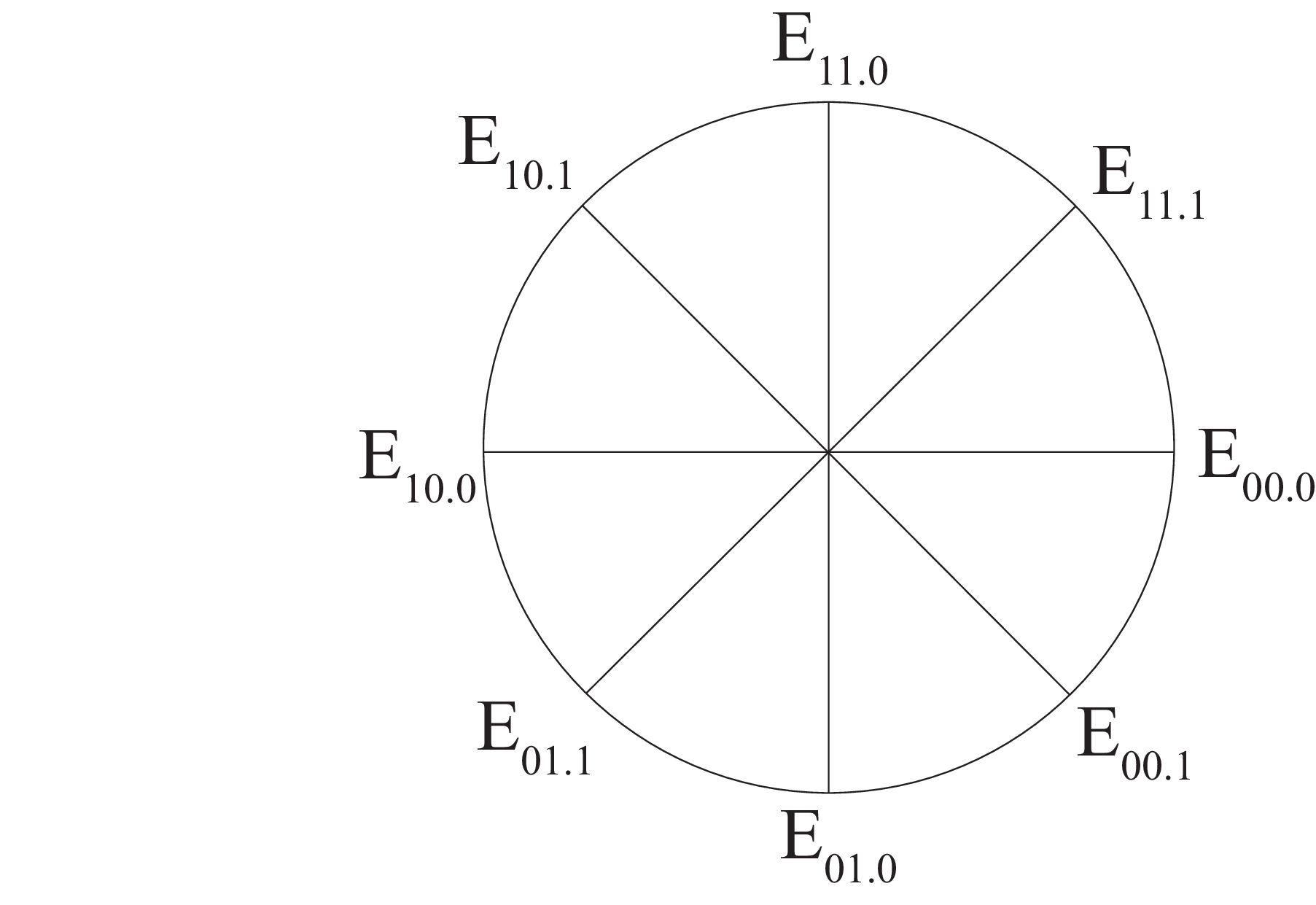}
\caption{A sequence of square-root-of-minus-one operators on the unit circle. Each pair of operators represented at points with angular separation $\pi/2$ are components of a quaternionic triple.}
\end{figure}

In terms of the Pauli permutation/negation operators, the Dirac permutation/negation operators are straightforwardly defined as
 \be
 \label{dirac}
\gamma_i= \bp \: &\sigma_i \\-\sigma_i& \; \ep
 \ee
suggesting, once the links to the complex Hilbert Space are developed below, a natural relationship with relativistic quantum theory.  

\subsubsection{Dyadic Rational Roots of Minus One}
\label{rootsofminusone}

Recall from (\ref{bar}) that $\mathbf{\bar E}_\beta$ is a $2^N \times 2^N$ block diagonal matrix containing $2^N/N$ copies of $\mathbf{E}_\beta$. As discussed above
\be
\mathbf{\bar E}^2_\beta=-1; \;\;\; \mathbf{\bar E}^3_\beta=-\mathbf{\bar E}_\beta; \;\;\; \mathbf{\bar E}^4_\beta=1
\ee
However, it is also possible to define fractional powers of $\mathbf{\bar E}_\beta$. Using the fact for any matrix $A$,  
\be
\bp A & \; \\ \; & A \ep =
\bp \; & 1 \\ A \; \ep \bp \; & 1 \\ A \; \ep  \nonumber
\ee
and the notation defined in (\ref{bar}), we can define 
\be
\label{example2}
 \mathbf{\bar{E}}_{\beta}^{1/2}= \overline{
\bp \;& 1 \\ 
\mathbf{E}_\beta &\; \ep}; \;\;
\mathbf{\bar{E}}_{\beta}^{1/4}= \overline{
\bp \;& \;& 1& \; \\
\;&\;&\;&1\\
\;&1&\;&\;\\
\mathbf{E}_\beta &\;&\;&\;\ep}; \;\; \ldots
\ee
Now the block element
\be
\bp \;& 1 \\ 
\mathbf{E}_\beta &\; \ep \nonumber
\ee
of  $\mathbf{\bar{E}}_\beta^{1/2}$ has size $2N \times 2N$. Similarly, the block element
\be
 \bp \;& \;& 1& \; \\
\;&\;&\;&1\\
\;&1&\;&\;\\
\mathbf{E}_\beta &\;&\;&\;\ep \nonumber
\ee
of $\mathbf{\bar{E}}_\beta^{1/4}$ has size $4N \times 4N$. Continuing in this way, the block element of $\mathbf{\bar{E}}_\beta^{1/2^N}$ has size $2^N \times 2^N$. This means that $\alpha=2^{-N}$ is the smallest exponent that can be defined,  given that $\mathbf{\bar{E}}_\beta^\alpha $ has size $2^N \times 2^N$. More generally, operators $\mathbf{\bar{E}}_\beta ^{\alpha}$ are defined provided $0 \le \alpha < 4$ where $\alpha \in \mathbb{Q}_2(N+2)$. If $\alpha \notin \mathbb{Q}_2(N+2)$, then $\mathbf{\bar{E}}_\beta ^{\alpha}$ is undefined. 

It is easily shown that the frequency of occurrence of the `$a$' symbol in the co-sequence
\be
|a')=\mathbf{\bar{E}}_\beta^{\alpha}\;\; |a) \nonumber
\ee
is, for all $\beta$, given by
\be
\label{prob}
P_a(\alpha)=|1-\frac{\alpha}{2}|
\ee
which can be interpreted as the probability of drawing the symbol `$a$' from $|a')$. For example, from (\ref{example2}), the probability of drawing an `$a$' from $|a')$ when $\alpha=1/4$ is (reading the matrix $\mathbf{\bar{E}}_\beta^{1/4}$ from top right to bottom left) 1/2+1/4+1/8=7/8. 

\subsubsection{Unitary Permutation/Negation Operators}
By analogy with complex-number matrices, the Hermitian transposes of the square-root-of-minus-one operators can readily be defined. Hence, returning to equation (\ref{three}) define 
 \be
\label{threestar}
\mathbf{E^*_0}=\bp -i&\;\\\; &i\ep;
\;\;\mathbf{E^*_1}=\bp \;&-i\\-i&\;\ep;
\;\;\mathbf{E^*_2}=\bp \;& 1\\ -1 &\; \ep \nonumber
\ee 
so that
\be
\mathbf{E}^*_0 \mathbf{E}_0=\mathbf{E}^*_1 \mathbf{E}_1=\mathbf{E}^*_2 \mathbf{E}_2=1 \nonumber
\ee
More generally, following the normal rules for the complex transpose
\be
\mathbf{E}^{*\alpha}_\beta \mathbf{E}^{\alpha}_\beta=1 \nonumber
\ee
Hence, if $|a')= \mathbf{\bar{E}}_{\beta}^{\alpha} \;|a) ;\;\;
|a'')=\mathbf{\bar{E}}_{\beta}^{\alpha\prime} \;|a)$ then $|a'')=\mathcal{\bar{U}}|a')$ where
\be
\label{unitary}
\mathcal{\bar{U}} = \mathbf{\bar{E}}_{\beta}^{\alpha\prime}\mathbf{\bar{E}}_{\beta}^{*\alpha}\ee
satisfies $\mathcal{\bar{U}}^*\mathcal{\bar{U}}=1$ and is therefore unitary.

\subsection{Directions on the Celestial Sphere}

In order to be able to make a connection to the world of experiment, it is fundamental that directions on the celestial sphere in physical 3-space can be described using the lbit ansatz above. The fact that the square-root-of-minus-one operators $\mathbf{\bar{E}}_\beta$ are quaternionic suggests a primitive relationship to rotations in physical 3-space may exist. 

Consider the 1-lbit state which is written in the form
\be
|a')_\beta^{\alpha}=\mathbf{\bar{E}}_\beta^{\alpha}\; |a) \nonumber
\ee
where, as before, $\beta \in \mathbb{Q}_2(\log_2 N +1)$, $\alpha \in \mathbb{Q}_2(N+2)$. 

Fig 4 shows how orientations on the celestial sphere in physical 3-space can be defined purely in terms of $|a')_\beta^{\alpha}$. Fig 3 shows a 2-sphere of unit radius. Inscribed within this sphere is an extremely thin cylinder (not shown to scale) whose height equals the diameter of the sphere and whose circumference is $N/2^N$ which is $\ll 1$ when $N \gg 0$. This cylinder can be considered a two dimensional symbolic representation of an `odd' iterate of $C$ associated with the symbolic label $a$. The `long-thinness' of the cylinder reflects the fact that $\alpha$ (representing the height of the cylinder) varies over $2^N$ values, whilst $\beta$ (representing the cylinder's angular coordinate) only varies over $N$ values. 

 \begin{figure}
\includegraphics[scale=0.7] {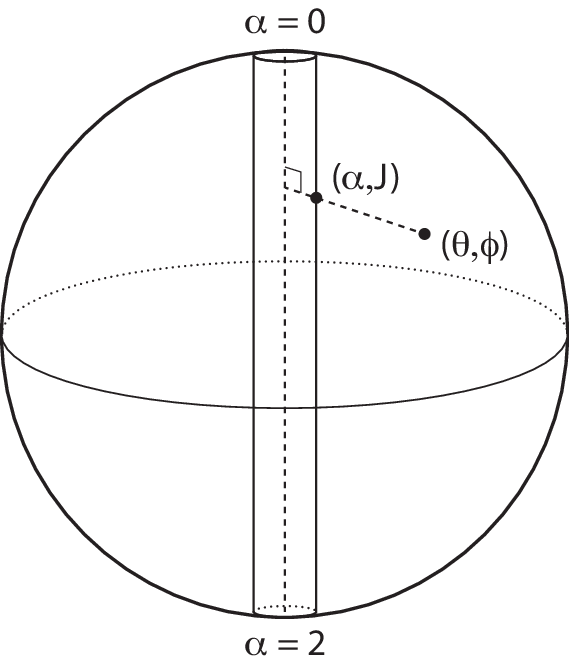}
\caption{A small cylinder whose circumference  to height ratio is $N:2^N$ is inscribed within a 2-sphere as shown. A point $(\alpha, \beta)$ on the cylinder corresponds to a co-sequence which describes the symbolic labelling of an element of one of the odd iterates of $C$. The illustrated mapping taking $(\alpha, \beta)$ to a point $(\theta, \phi)$ on the sphere, where $\theta$ denotes latitude relative to the North Pole and $\phi$ longitude, shows how the celestial sphere of orientations in physical 3-space is related to the symbolic representation of $C$.}
\end{figure}

Now $\alpha$ describes the distance around the cylinder `in the long direction', i.e. going from the top to the bottom and back up to the top again. As  $\alpha$ varies from 0 to 4, the frequency of occurrence of the `$a$' symbol in $|a')_\beta^{\alpha}$ varies as
\be
P_a(\alpha)=|1-\frac{\alpha}{2}|
\ee
Hence, for example, if $\alpha=0$, all elements of $|a')_\beta^{\alpha}$ are `$a$'s and if $\alpha=2$, all elements of $|a')_\beta^{\alpha}$ are `$\lnot a$'s. Hence $\alpha=0$ and $\alpha=2$ lie at the top and bottom of the cylinder respectively.  If $\alpha=1$ or $\alpha=3$, half the members of $|a')_\beta^{\alpha}$ are `$a$'s, the rest are `$\lnot a$'s and these are antipodal points on the `equatorial' circle of the cylinder. 

Now take an arbitrary point on the cylinder associated with $|a')_\beta^{\alpha}$ and project the line orthogonal to the axis of the cylinder from the centre of the cylinder and through $(\alpha, \beta)$ until it reaches the sphere (see Fig 4). We define this point on the sphere, labelled by $(\theta, \phi)$, where $\theta$ is latitude and $\phi$ longitude, as an orientation in 3-space. That is to say, the construction above defines a point on the celestial sphere from the primitive lbit parameters $(\alpha, \beta)$. Given that $N \ggg 0$ (i.e. the radius of the cylinder can be treated as if infinitesimally small compared with its height), the relationship between $\theta$ and $\alpha$ is 
\be
\label{nature}
\cos^2 \frac{\theta}{2}= | 1 - \frac{\alpha}{2} |
\ee
or equivalently
\begin{align}
\label{nature2}
\alpha&=1-\cos \theta\;\;\;\mathrm{if}\;\;\;0\le\theta \le \pi \nonumber\\
\alpha&=3+\cos\theta\;\;\;\mathrm{if}\;\;\;\pi \le \theta \le 2\pi
\end{align}
Hence if $\alpha \in \mathbb{Q}_2(N+1)$ (see Section \ref{rootsofminusone}), then so is $\cos \theta$. 

The corresponding relationship between $\beta$ and the longitudinal coordinate $\phi$ on the celestial sphere is simply given by 
\be
\label{nature3}
\phi= \frac{\pi}{2} \; \beta
\ee
Since $\beta$ is a dyadic rational, then the angles $\phi$ are dyadic rational fractions of $\pi$. As discussed in the appendix, except for $0, \pi/2, \pi, 3\pi/2$, the cosine of such angles are irrational numbers and hence do not belong to the set $\mathbb{Q}_2(N+2)$. This is fundamental to all that follows. 

\section {IST and Quantum Theory}
\label{quantum}

The formalism above in applied to the quantum physics of multi-qubits and in so doing a new interpretation of (the complex Hilbert Space in) quantum theory is uncovered. 

\subsection{IST and Counterfactual Incompleteness}
\label{counterfactual}

To show the crucial difference between IST and quantum theory, consider the following example. From a quantum theoretic point of view, the points $p'_1$ and $p'_2$ in Fig 5a, considered as a Bloch Sphere, represent two non-entangled qubit states (e.g. representing spin states of two spin-1/2 particles), each a superposed state relative to a basis oriented in the direction $\mathbf{\hat{z}}$ represented by the North Pole. If the eigenstates at the North/South poles are referred to as `$|a\rangle$' and `$|\lnot a \rangle$', then we can write
\begin{align}
p^\prime_1 \sim |a\rangle &+ |\lnot a\rangle \nonumber \\
p^\prime_2 \sim |a\rangle &+ e^{i\phi} |\lnot a\rangle \nonumber
\end{align}
respectively, where $0 < \phi < \pi/2$. Suppose a spin-1/2 system is prepared in the North Pole `up' state, then there is a well defined probability of being measured `up' in either the $p'_1$ or $p'_2$ direction which equals 1/2. Counterfactually, if the spin-1/2 system had been prepared in the 'up' direction relative to the $p'_1$ direction, then the probability of being measure 'up' in the $p'_1$ or $p'_2$ direction would have been equal to $1$ and $\cos^2 \phi/2$ respectively. 

Let us now study this example from the perspective of IST using the mapping defined in Fig 4. We can write 
\begin{align}
\label{qi}
p^\prime_1 \sim \mathbf{\bar{E}}_{0}|a) \nonumber \\
p^\prime_2 \sim  \mathbf{\bar{E}}_{\beta}|a) 
\end{align}
where the north pole represents the sequence $|a)$ comprising entirely of `$a$'s. From (\ref{nature3}), $\phi= \pi\beta/2$. 

Again, if the system is prepared in the North Pole `up' state, then there is a well defined probability of being measured `up' in either the $p'_1$ or $p'_2$ direction will equal 1/2. Counterfactually, if the spin-1/2 system had been prepared in the 'up' direction relative to the $p'_1$ direction, then the probability of being measure 'up' in the $p'_2$ direction would be undefined. To see this, note from the simple number-theoretic argument in the Appendix, if $0<\phi< \pi/2$ is a dyadic rational multiple of $\pi$ then $\cos^2 \phi/2$ is certainly irrational. 

\begin{figure}
\includegraphics[scale=0.7] {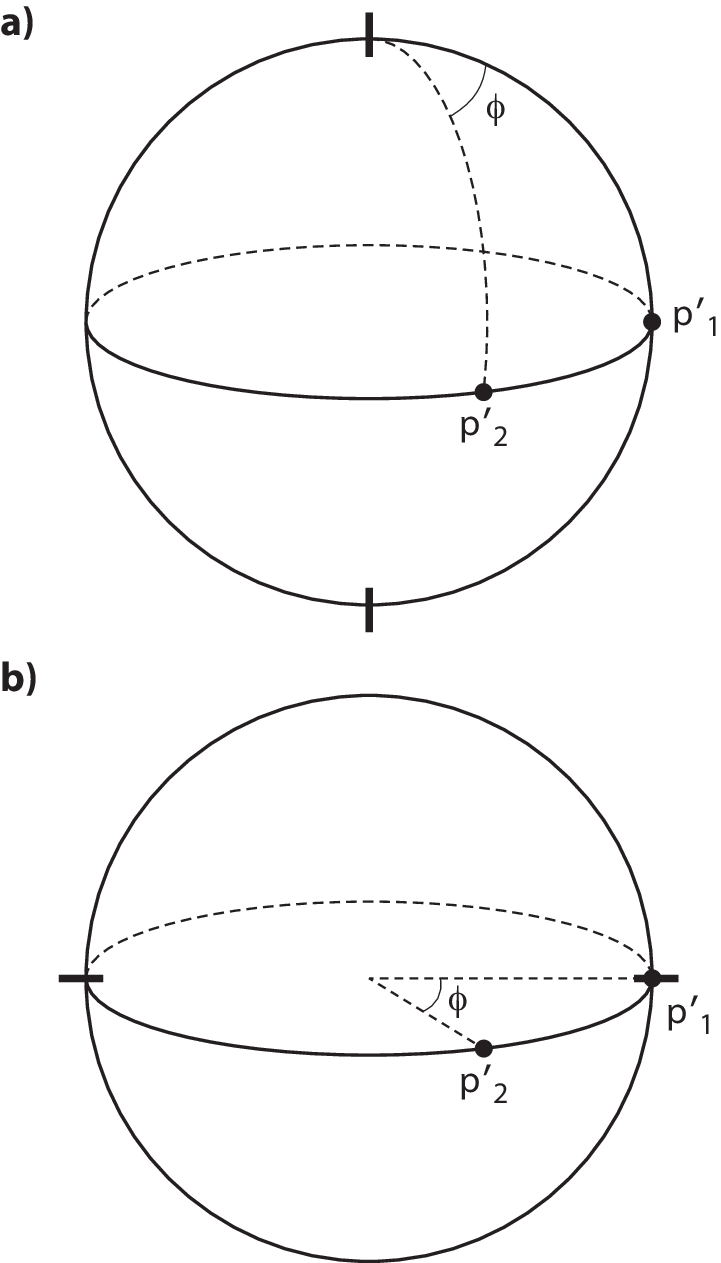}
\caption{a) A qubit spin state is prepared `up' in the $\mathbf{\hat{z}}$ direction and measured in either the $p'_1$ or $p'_2$ direction. From an IST perspective, two lbit states are associated with co-sequences of symbolic labels, such that $0<\phi< \pi/2$ is a dyadic rational multiple of $\pi$. b) From a quantum theoretic perspective if the qubit had been prepared `up' in the $\mathbf{\hat{x}}$ direction, the probability of being measured `up' in the $p'_2$ direction is $\cos^2 \phi/2$. From an IST perspective, if the system had been prepared as spin `up' in the $\mathbf{\hat{x}}$,  then the probability of being measured `up' in the $p'_2$ direction is undefined because $\cos\phi$ must be irrational. This example demonstrates a generic counterfactual incompleteness associated with the sparseness of $\mathcal{I}_D$ in state space, and illustrates why IST is a contextual theory.}
\end{figure}

Recall that two key attributes of $\mathcal{I}_D$ are firstly self similarity and secondly sparseness in the Euclidean space in which $\mathcal{I}_D$ can be embedded. The fact that $\mathcal{I}_D$ has measure zero implies that if a point on $\mathcal{I}_D$ is somehow perturbed randomly with respect to the measure of this embedding space, then the perturbed point will almost certainty not lie on $\mathcal{I}_D$. If the system was prepared in the $\mathbf{\hat{z}}$ direction, then the perturbation which reorientates the preparation apparatus to the $\mathbf{\hat{x}}$, keeping the particles unchanged, takes the space-time off the invariant set. That is to say, the relationships (\ref{nature}) (\ref{nature2}) and (\ref{nature3}) are consistent with the the sparseness of $\mathcal{I}_D$. One could indeed view these relationships as being required by the sparseness of $\mathcal{I}_D$. Hence IST can describe the experimental quantum physics of qubits, but in a necessarily contextual fashion (the co-sequence space is linked to an orientation in physical space). By contrast, the Hilbert Space of quantum theory is non-contextual (the Hilbert Space is covariant and not linked to any particular choice of measurement orientation). In quantum theory, it is an irrelevance whether a particular measurement is counterfactual or not. 

This gives us a novel perspective on the Hilbert Space, which is developed in the next section. 

\subsection{Symbolic Dynamics and the Complex Hilbert Space}

The relationship between IST and quantum theory can begin to be addressed, and this gives new insights into why the complex Hilbert Space representation of state plays such a central role in quantum physics. 

If the mutually exclusive events `$a$' and `$\lnot a$' have probability $p(a)$ and $p(\lnot a)=1-p(a)$, then if $\mathbf{j}_a$ and $\mathbf{j}_{\lnot a}$ are unit vectors in $\mathbb{R}^2$, Pythagoras' Theorem implies that
\be
\label{hilbert}
\mathbf{v}=\sqrt{p(a)} \;\; \mathbf{j}_a + \sqrt{p(\lnot a)} \;\; \mathbf{j}_{\lnot a}
\ee
is itself a unit vector, and can therefore represent the frequentist-based probability of drawing an $a$ from $|a')$. 

With this in mind, consider the correspondence
\be
\label{bla}
\mathbf{\bar{E}}_{0.00 \ldots}^{\alpha}\;|a) \sim \cos \frac{\theta}{2} |a\rangle + \sin \frac {\theta}{2} |\lnot a\rangle 
\ee
where, repeating (\ref{nature}) 
\be
\cos^2 \frac{\theta}{2} = | 1 - \frac{\alpha}{2}|
\ee
The symbol `$\sim$' in (\ref{bla}) should be interpreted as follows: if $X_a \sim x_a$ then the probability of drawing an $a$ from the co-sequence $X_a$ can be represented by the Hilbert Space element $x_a$. As ever, $\cos \theta \in \mathbb{Q}_2$. Recalling that $\mathbf{\bar{E}}_{0.00\ldots}$ and $\mathbf{\bar{E}}_{1.00\ldots}$ are quaternionic pairs, we can supplement (\ref{bla}) with
\begin{align}
\label{qubit}
  \mathbf{\bar{E}}_{1.00 \ldots}^{\alpha}\;|a) &\sim \cos \frac{\theta}{2} |a\rangle + i \sin \frac {\theta}{2} |\lnot a\rangle  
\end{align}
Finally, for a general value $\beta$ put 
\be
\mathbf{\bar{E}}_{\beta}^{\alpha} |a) \sim \cos \frac{\theta}{2} |a\rangle + i^{\beta} \sin \frac {\theta}{2} |\lnot a\rangle 
\ee
or
\be
\label{qubit0}
\mathbf{\bar{E}}_{\beta}^{\alpha}|a) \sim \cos \frac{\theta}{2} |a\rangle + e^{i\phi} \sin \frac {\theta}{2} |\lnot a\rangle 
\ee
where $\phi= \pi\beta/2$. 

The notion of probability has been used to relate \emph{all} the co-sequences needed in IST, to complex Hilbert Space states. However, crucially, not all Hilbert Space states correspond to co-sequences. As described in Section \ref{counterfactual}, this is related to the fact that the invariant set is `gappy'. 

However, Pythagoras' Theorem allows us to `fill in the gaps' since it is possible to write
 \be
\label{hilbert2}
\mathbf{v}=\cos \frac{\theta}{2 }\;\; \mathbf{j}_a + \sin \frac{\theta}{2} \;\; \mathbf{j}_{\lnot a}
\ee
irrespective of whether $\cos^2 \theta/2$, and hence $\cos \theta$, is rational. By continuing $\cos \theta$ to the irrationals, we can extend the frequentist probabilistic representation of state to form a continuum vector space. It should be recognised that the elements of the vector space associated with irrational values of $\cos \theta$ are essentially algebraically abstract quantities. 

In physics, we are accustomed to viewing closed algebraic structures as `good things' on which to base our physical theory. However, this may be a mistaken perspective. Certainly the Hilbert Space, as a closed vector space, has enormous computational advantages over the symbolic sequences of IST, however, as is well known, the Hilbert Space is paradoxical - it predicts half dead, half alive cats!

There is an analogy here with the real numbers themselves. There is no doubt that the concept of real numbers is mathematically beautiful, and moreover they are a calculational convenience for physicists. (Who would want to do calculus on the rationals?) However, unfettered use of the real numbers in physics can lead to paradoxes: the Banach-Tarski theorem, which uses non-measurable subsets of the sphere, makes a mockery of physical conservation laws, for example. Here we would argue that just as the real numbers shouldn't be taken too seriously in physical theory, neither should the Hilbert Space. 

By continuing both $\cos \theta$ and $\pi\beta=2\phi$ to the continuum, the resulting vector space is the complex Hilbert Space of quantum theory. 

\subsection{Unitary Evolution}

The sequential SG apparatus provides a paradigmatic example of state evolution in quantum theory. A system is prepared in a certain state and then measured. This measurement in turn   can be considered as the means of preparing the system for a second measurement, and so on. In quantum-theoretic language, the preparation basis and the measurement basis are not in general the same, and the transformation of the state of the system between the preparation basis and the measurement basis is associated with a unitary operator. In conventional quantum theory, the measurement process `collapses' the generically superposed state to one of the eigenstates of the measurement basis. This eigenstate defines the preparation state associated with the next of the sequential measurements. In the standard Copenhagen interpretation, the collapse process is not considered unitary, and in theories where the collapse process is modelled explicitly, unitarity is manifestly violated. However, modern decoherence theory provides a model for measurement which on the one hand is unitary, but on the other hand implies that the (baffling) physical concept of state superposition is not restricted to the micro-scale. 

The interpretation of state evolution in such sequential measurement situations is quite different in IST and this provides a radically new prospective on the notion of unitarity. We start with a set of co-sequences $\mathbf{E}^\alpha_\beta |a)$ defined relative to some arbitrary orientation $\mathbf{\hat{z}}$ on the celestial sphere (see e.g. Fig 6a). For each (dyadic rational) $\alpha$, $\beta$, these define a sample space of bivalent outcomes (`$a$' or `$\lnot a$') for a set of measurement orientations (countably infinite in the limit $N \rightarrow \infty$) relative to the preparation direction $\mathbf{\hat{z}}$. Let us fix on one of these orientations $\mathbf{\hat{z^\prime}}$. The co-sequence associated with this orientation defines the symbolic representation of some iterate $C_{i+1}$ of the invariant set. 

Relative to $\mathbf{\hat{z^\prime}}$ we have a second set of co-sequences $\mathbf{E}^\alpha_\beta |b)$ on the celestial sphere, defining a second space of bivalent outcomes (`$b$' or `$\lnot b$') for a set of measurement orientations relative to $\mathbf{\hat{z^\prime}}$ (see e.g. Fig 6b). Let us fix on one of these orientations $\mathbf{\hat{z^{\prime\prime}}}$. The co-sequence associated with this orientation defines the symbolic representation of the next odd iterate $C_{i+3}$ of the invariant set. In this way, each odd iterate $C_{i + 2n+1}$ has a symbolic representation. As discussed above, all the permutation/negation operators $\mathbf{E}^\alpha_\beta$ are unitary. Hence the sequence of preparations and measurements can be considered unitary in IST. In this sense IST seems closer to decoherence theory, than theories which have collapse models. 

On the other hand, as discussed above, unitarity fails in IST when we consider a transformation from a prepared state to a counterfactual measurement state. To see this, return to the $\mathbf{E}^\alpha_\beta |a)$ defined relative to the first arbitrary preparation orientation $\mathbf{\hat{z}}$, choose one of the co-sequences, hence defining the first measurement orientation $\mathbf{\hat{z^\prime}}$. Now ask what would be the sample space when the state is prepared relative to $\mathbf{\hat{z^\prime}}$ but measured with respect to one of the directions $\mathbf{\hat{z_c^\prime}}$ that might have been chosen but wasn't. As discussed above, there is no sample space, even in the limit $N \rightarrow \infty$ when the set of possible orientations was countably infinite. 

Table 1 shows the key difference between quantum theory (with and without collapse models) and IST. 

\vspace{7 mm}
\begin{center}
\begin{tabular}{|c|c|c|} \hline
quantum theory                           & time evolution & unitary \\ \cline{2-3}
with decoherence                & counterfactual $\textrm{X}^n$& unitary \\ \hline
quantum theory                & time evolution & non-unitary \\ \cline{2-3}
with collapse                     & counterfactual $\textrm{X}^n$& unitary \\ \hline
invariant set          & time evolution & unitary \\ \cline{2-3}
theory                                         & counterfactual $\textrm{X}^n$& non-unitary \\ \hline
\end{tabular}
\\
\vspace{2mm}
TABLE 1
\end{center}

\vspace{5 mm}                      

\subsection{Entanglement}
 \label{EI}
 
Consistent with the discussion earlier in this paper, $\mathcal{I}_D$ is to be considered a multi-dimensional Cantor Set $C$ of state-space trajectories, each trajectory defining a space-time. The symbolic representation of $\mathcal{I}_D$ is defined by the two-parameter set
\be
\{\mathbf{\bar{E}}_\beta^{\alpha}\} \nonumber
\ee
of square-root-of-minus-one operators defined above. Each co-sequence $|a')_\beta^{\alpha}=\mathbf{\bar{E}}^\alpha_\beta |a)$ labels a grouping of elements of one of the iterates $C_2, C_4, C_6 \ldots$ of $C$, i.e. provides a symbolic description of one of the elements of $C_1, C_3, C_5 \ldots$. 

We now consider a multi-symbolic labelling of $C$. Let $a_1, a_2 \dots a_{n}$ denote $n$ symbols which label symbolically a trajectory of $\mathcal{I}_D$.  Recall from Section \ref{sequential} that the symbolic representation of the iterates of $C$ are based on sequences or co-sequences of symbols associated with the labels of trajectories through these iterates. These co-sequences will in turn be based on a product of quaternion operators acting on $|a_1), |a_2), \ldots |a_{n})$ respectively. That is to say, a general form for the symbolic representation of the iterates of $C$ will be given by  
\begin{align}
\label{nlbit}
|a'_1)=& (\prod_{j=1}^{n_{\mathrm{col}}} \mathcal{P}_{1j}) |a_1) \nonumber \\
|a'_2)=& (\prod_{j=1}^{n_{\mathrm{col}}} \mathcal{P}_{2j}) |a_2) \nonumber \\
\ldots \nonumber \\
|a'_{n})=& (\prod_{j=1}^{n_{\mathrm{col}}} \mathcal{P}_{nj}) |a_{n}) \nonumber \\
\end{align}
where $\mathcal{P}_{ij}$ denotes an array of quaternion operators taken from $\{\mathbf{\bar{E}}^\alpha_\beta\}$ defined above. The representation in (\ref{nlbit}) is called an $n$ `labelling bit' or an `$n$-lbit' for short. As will be shown, the correspondence between $n$-lbits and $n$-qubits is striking. 

By definition, $\mathcal{P}_{ij}$ has $n$ rows (one for each label). The construction of $\mathcal{P}_{ij}$, and the determination of the number of columns $n_{\mathrm{col}}$ is based on a combinatoric ansatz, defined as follows. This combinatoric approach attempts to describe all possible linkages between the individual lbits. All $n$ entries in the last column of $\mathcal{P}_{ij}$ are the same quaternion operator. Working backwards from the last column to the first, each of the next $\binom{n}{n-1}$ columns of $\mathcal{P}_{ij}$ describe all permutations of at most two quaternion operators where $n-1$ entries are the same operator. The next $\binom{n}{n-2}=\binom{n}{2}$ columns of $\mathcal{P}_{ij}$ describe all permutations of at most two quaternion operators where $n-2$ entries are the same quaternionic operator and the remaining two rows are themselves the same operator. We continue in this way until we reach $\binom{n}{\lceil n/2 \rceil}$. Using the elementary relationship
\be
\sum_{k=\lceil n/2 \rceil}^{n} \binom{n}{k}=2^{n-1} \nonumber
\ee
then $n_{\mathrm{col}}=2^{n-1}$. Apart from the last column of $\mathcal{P}_{ij}$ (which contains just one operator), all columns contain at most two operators. Hence at most $2n_{\mathrm{col}}-1=2^n-1$ operators are used in describing $\mathcal{P}_{ij}$. 

For example, with $n=1$, $n_{\mathrm{col}}=1$ and 
\be
\label{1lbit}
|a')=\mathbf{\bar{E}}_{\beta_1}^{\alpha_1} \;|a)
\ee
where $J_1=1$. With $n=2$,  and 
\begin{align}
\label{2lbits}
|a')&=\mathbf{\bar{E}}_{\beta_1}^{\alpha_{1}}\mathbf{\bar{E}}_{\beta_3}^{\alpha_{3}}\;|a) \nonumber \\
|b')&=\mathbf{\bar{E}}_{\beta_2}^{\alpha_{2}}\mathbf{\bar{E}}_{\beta_3}^{\alpha_{3}}\;|b)
\end{align}
With $n=3$ there are  and
\begin{align}
\label{3lbits}
|a')&=\mathbf{\bar{E}}_{\beta_1}^{\alpha_1}\mathbf{\bar{E}}_{\beta_3}^{\alpha_3}\mathbf{\bar{E}}_{\beta_5}^{\alpha_5}\mathbf{\bar{E}}_{\beta_7}^{\alpha_7}\;|a) \nonumber \\
|b')&=\mathbf{\bar{E}}_{\beta_2}^{\alpha_2}\mathbf{\bar{E}}_{\beta_3}^{\alpha_3}\mathbf{\bar{E}}_{\beta_6}^{\alpha_6}\mathbf{\bar{E}}_{\beta_7}^{\alpha_7}\;|b) \nonumber \\
|c')&=\mathbf{\bar{E}}_{\beta_2}^{\alpha_2}\mathbf{\bar{E}}_{\beta_4}^{\alpha_4}\mathbf{\bar{E}}_{\beta_5}^{\alpha_5}\mathbf{\bar{E}}_{\beta_7}^{\alpha_7}\;|c)
\end{align}
With $n=4$ $n_{\mathrm{col}}=8$ and
\begin{align}
\label{4lbits}
|a')&=\mathbf{\bar{E}}_{\beta_1}^{\alpha_1}\mathbf{\bar{E}}_{\beta_3}^{\alpha_3}\mathbf{\bar{E}}_{\beta_5}^{\alpha_5}\mathbf{\bar{E}}_{\beta_7}^{\alpha_7}\mathbf{\bar{E}}_{\beta_9}^{\alpha_9}\mathbf{\bar{E}}_{\beta_{11}}^{\alpha_{11}}\mathbf{\bar{E}}_{\beta_{13}}^{\alpha_{13}}\mathbf{\bar{E}}_{\beta_{15}}^{\alpha_{15}}\;\;|a) \nonumber \\
|b')&=\mathbf{\bar{E}}_{\beta_2}^{\alpha_2}\mathbf{\bar{E}}_{\beta_3}^{\alpha_3}\mathbf{\bar{E}}_{\beta_6}^{\alpha_6}\mathbf{\bar{E}}_{\beta_7}^{\alpha_7}\mathbf{\bar{E}}_{\beta_9}^{\alpha_9}\mathbf{\bar{E}}_{\beta_{12}}^{\alpha_{12}}\mathbf{\bar{E}}_{\beta_{14}}^{\alpha_{14}}\mathbf{\bar{E}}_{\beta_{15}}^{\alpha_{15}}\;\;|b) \nonumber \\
|c')&=\mathbf{\bar{E}}_{\beta_2}^{\alpha_2}\mathbf{\bar{E}}_{\beta_4}^{\alpha_4}\mathbf{\bar{E}}_{\beta_5}^{\alpha_5}\mathbf{\bar{E}}_{\beta_7}^{\alpha_7}\mathbf{\bar{E}}_{\beta_{10}}^{\alpha_{10}}\mathbf{\bar{E}}_{\beta_{11}}^{\alpha_{11}}\mathbf{\bar{E}}_{\beta_{14}}^{\alpha_{14}}\mathbf{\bar{E}}_{\beta_{15}}^{\alpha_{15}}\;|c) \nonumber \\
|d')&=\mathbf{\bar{E}}_{\beta_1}^{\alpha_1}\mathbf{\bar{E}}_{\beta_4}^{\alpha_4}\mathbf{\bar{E}}_{\beta_6}^{\alpha_6}\mathbf{\bar{E}}_{\beta_8}^{\alpha_8}\mathbf{\bar{E}}_{\beta_9}^{\alpha_9}\mathbf{\bar{E}}_{\beta_{11}}^{\alpha_{11}}\mathbf{\bar{E}}_{\beta_{14}}^{\alpha_{14}}\mathbf{\bar{E}}_{\beta_{15}}^{\alpha_{15}}\;\;|d)
\end{align}
The procedure can be continued to arbitrarily large $n$ (though to write down the explicit formulae will require the development of a more compact notation!). 

Conversely, within (\ref{4lbits}) there are sub-arrays which have the structure of the arrays $\mathcal{P}_{ij}$ for 1-lbits, 2-lbits, 3-lbits. For example, the top left $1 \times 1$, $2 \times 2$ and $3 \times 4$ sub-arrays of operators in (\ref{4lbits}) have the structure of the operator arrays for 1, 2 and 3 -lbits in (\ref{1lbit}), (\ref{2lbits}) and (\ref{3lbits}) respectively. 

In summary, a 1-lbit has 2 free parameters: $\alpha_1$ and $\beta_1$. A 2-lbit has 6 free parameters:  $\alpha_1, \alpha_2, \alpha_3$ and $\beta_1, \beta_2, \beta_3$. A 3-lbit has 14 free parameters:  $\alpha_1, \alpha_2, \ldots, \alpha_7$ and $\beta_1, \ldots \beta_7$. In general, an $n$- lbit has $2^{n+1}-2$ free parameters. This is same as the dimension of the complex Hilbert Space of $n$ qubits in quantum theory, factoring out the normalisation and global phase. 

The multi-labelling ansatz described in this sub-section is the basis for entanglement correlation in invariant set theory. For example, the expression for a 2-lbit state (cf (\ref{2lbits})) can be written as: 
\be
|a')=\mathbf{\bar{E}}_{\beta_1}^{\alpha_1}\mathbf{\bar{E}}_{\beta_3}^{\alpha_3}\;|a); \;\;\;
|b')=\mathbf{\bar{E}}_{\beta_2}^{\alpha_2}\mathbf{\bar{E}}_{\beta_3}^{\alpha_3}\;|b) \nonumber
\ee
Putting $\beta_2=\beta_1$ and writing $a \equiv a_{\mathrm{Alice}}$ and $b \equiv a_{\mathrm{Bob}}$, then  
\begin{align}
|a'_{\mathrm{Alice}})&=\;\;\;\;\;\;\;\;\;\;\;\mathbf{\bar{E}}_{\beta_1}^{\alpha_1}\mathbf{\bar{E}}_{\beta_3}^{\alpha_3}\;|a_{\mathrm{Alice}}) \nonumber \\
|a'_{\mathrm{Bob}})\;&=\mathbf{\bar{E}}_{\beta_1}^{\alpha_2-\alpha_1} \;\mathbf{\bar{E}}_{\beta_1}^{\alpha_1}  \mathbf{\bar{E}}_{\beta_3}^{\alpha_3}\;|a_{\mathrm{Bob}}) \nonumber
\end{align}
Hence the correlation between the  $|a'_{\mathrm{Alice}})$ and $|a'_{\mathrm{Bob}})$ co-sequences is determined by the exponent $\alpha_2 - \alpha_1$ in the operator $\mathbf{\bar{E}}_{\beta_1}^{\alpha_2-\alpha_1}$. Just as the probability of drawing an `$a$' symbol in $\mathbf{\bar{E}}_{\beta_1}^{\alpha_2-\alpha_1}|a)$ is $|1-\frac{\alpha_1-\alpha_2}{2}|$ so too is the probability that an element of $|a'_{\mathrm{Alice}})$ agrees with a corresponding element of  $|a'_{\mathrm{Bob}})$. Using the relationship defined in Fig 3 between elements of the symbolic representation of $\mathcal{I}_D$ and points on the celestial sphere, so that   
\be
\label{EPR}
\cos^2 \frac{\theta}{2} = |1-\frac{\alpha_2-\alpha_1}{2}|
\ee
then IST describes the quantum mechanical correlation $\cos \theta$ between Alice and Bob's measurements on the entangled state
\be
\frac{|\uparrow\rangle|\uparrow\rangle-|\downarrow\rangle|\downarrow\rangle}{\sqrt{2}} \nonumber
\ee
where $\theta$ now defines the relative orientation of Bob and Alice measurement apparatuses.  

A key question, of course, is whether IST is constrained by Bell inequalities. As before, we require $\cos \theta \in \mathbb{Q}_2(N)$ but this is hardly a constraint on the possible orientations available (for all practical purposes) to Alice and Bob. Indeed, suppose $\cos \theta' \in \mathbb{Q}_2(N)$ denotes another possible relative orientation for Alice and Bob, so that
\be
\cos^2 \frac{\theta'}{2} = |1-\frac{\alpha'_2-\alpha'_1}{2}| \nonumber
\ee
and $\alpha'_1, \alpha'_2 \in \mathbb{Q}_2(N)$. Now Bell's inequality states that for standard local hidden-variable theories
\be
\label{Bell}
|C(\theta)-C(\theta')|-C(\theta - \theta') \le 1
\ee
where $C$ is a hidden-variable correlation. Is IST constrained by this inequality? To be constrained by (\ref{Bell}), each of the three correlations $C(\theta)$, $C(\theta')$ and $C(\theta-\theta')$ must be well defined. But in IST, the third is not if the first two are well defined. The reason is the same as that discussed in Section \ref{counterfactual}. 
 
 It is well known that Bell's Theorem fails without counterfactual definiteness (see e.g. \cite{Bell}). The challenge has always been to develop a mathematical formalism consistent with quantum physics, in which such failure arises naturally and not in some contrived (or `conspiratorial') way. In IST the Bell Theorem fails because $\cos 2\theta$ is quadratic, and not linear, in $\cos\theta$: from Appendix A, this is the primary number-theoretic origin of why the symbolic representation of $\mathcal{I}_D$ exhibits counterfactual incompleteness. One could hardly imagine a less contrived and less conspiratorial solution to this problem! At a more physical level, these number theoretic properties reflect, symbolically, the fact that $\mathcal{I}_D$ has measure zero in state space, so that counterfactual perturbations such as associated with the above, take points on $\mathcal{I}_D$, off $\mathcal{I}_D$. Of course, the very postulate that states of reality lie on $\mathcal{I}_D$ is itself a `global' condition. However, this postulate is not `nonlocal', if by nonlocality we mean`not locally causal'. Rather the invariant set postulate should be thought of as a mathematical embodiment of the Bohmian notion of an `undivided universe'. Hence IST, although realistic, is not non-local, i.e. the relativistic principle of local causality does not fail. This is consistent with the speculation (see below) that IST extends general relativity by being geometric in both space time and state space. 

In passing, the formalism in Section \ref{quaternion} allows a description of entanglement for multiple qubit states. For example, with $\beta_1=\beta_2=\beta_3=\beta_4=\beta_5=\beta_6$, (\ref{3lbits}) becomes
\begin{align}
|a')&=\mathbf{\bar{E}}_{\beta_1}^{\alpha_1}\mathbf{\bar{E}}_{\beta_1}^{\alpha_3}\mathbf{\bar{E}}_{\beta_1}^{\alpha_5}\mathbf{\bar{E}}_{\beta_7}^{\alpha_7}\;|a) \nonumber \\
|b')&=\mathbf{\bar{E}}_{\beta_1}^{\alpha_2}\mathbf{\bar{E}}_{\beta_1}^{\alpha_3}\mathbf{\bar{E}}_{\beta_1}^{\alpha_6}\mathbf{\bar{E}}_{\beta_7}^{\alpha_7}\;|b) \nonumber \\
|c')&=\mathbf{\bar{E}}_{\beta_1}^{\alpha_2}\mathbf{\bar{E}}_{\beta_1}^{\alpha_4}\mathbf{\bar{E}}_{\beta_1}^{\alpha_5}\mathbf{\bar{E}}_{\beta_7}^{\alpha_7}\;|c) \nonumber
\end{align}
which in turn can be written
\begin{align}
|a')&=\mathbf{\bar{E}}_{\beta_1}^{\beta_1}\; \mathbf{\bar{E}}_{\beta_7}^{\alpha_7}\;|a) \nonumber \\
|b')&=\mathbf{\bar{E}}_{\beta_1}^{\beta_2} \;\mathbf{\bar{E}}_{\beta_7}^{\alpha_7}\;|b) \nonumber \\
|c')&=\mathbf{\bar{E}}_{\beta_1}^{\beta_3} \;\mathbf{\bar{E}}_{\beta_7}^{\alpha_7}\;|c) \nonumber
\end{align}
where $\beta_1=\alpha_1+\alpha_3+\alpha_5$, $\beta_2=\alpha_2+\alpha_3+\alpha_6$ and $\beta_3=\alpha_2+\alpha_4+\alpha_5$. These co-sequences describe the measurement statistics associated with the Hilbert Space state
\be
\frac{|\uparrow\rangle|\uparrow\rangle|\uparrow\rangle-|\downarrow\rangle|\downarrow\rangle|\downarrow\rangle}{\sqrt{2}} \nonumber
\ee
for the three measurement orientations
\be
\cos^2 \frac{\theta_1}{2}=| 1- \frac{\beta_1}{2}|; \;\cos^2 \frac{\theta_2}{2}=| 1- \frac{\beta_2}{2}|; \;\cos^2 \frac{\theta_3}{2}=| 1- \frac{\beta_3}{2}| \nonumber
\ee. 

\section{Towards a Gravitational Theory of the Quantum}
\label{GQ}
 
 For well over a half century, theoretical physicists have been actively seeking a quantum theory of gravity \cite{Rovelli} \cite{Maldacena:2004}. Not least, such a theory is generally considered necessary to explain the Big Bang and the singularities of black holes. Moreover, it is not considered logical to describe particles using quantum mechanics but space-time with classical physics. For example, since matter causes space-time to curve, then if it is possible to consider a particle which is in a quantum mechanical superposition of two states with different positions, so the gravitational field associated with this particle should be in a similar superposition. Of course, this is possible only if the gravitational field is quantised. In the two principal contenders for a quantum theory of gravity, string theory and loop quantum gravity,  this notion of a superposed gravitational field is considered primitive. 
 
 Here a completely different perspective has been presented, one for which the superposed state does not have any ontological significance and is rather a computational convenience in the light of the algorithmic intractability of $\mathcal{I}_D$. In particular, it has been suggested that quantum physics might be emergent from a generalised theory of gravity which retains and develops some of general relativity's key features: determinism, causality, geometry. That is to say, rather than seem a quantum theory of gravitation, it is suggested here that we should be looking for a gravitational theory of the quantum.  
 
Much of the motivation for the type of fractal state-space geometry discussed in this paper has come from the theory of forced dissipative multi-scale nonlinear systems (such as weather) which exhibit fractal invariant-set attractors. In such cases, the systems are maintained against small-scale dissipation by some large-scale forcing. In the current context it has been speculated that the invariant sets of cosmological space-times arise from a balance between forcing on the cosmological scale associated with a positive cosmological constant, and state space convergence of trajectories on the Planck scale. The phenomenon of information loss at black hole singularities \cite{Penrose:2011} is entirely consistent with state-space trajectory convergence. (And so, rather than invoke quantum theory as the key to understanding space-time singularities, maybe space-time singularities must be invoked as the key to understanding quantum theory!)
 
 Many physicists are uncomfortable with the notion of information loss. However, it is not directly relevant at the laboratory scale. A good analogy here is with the phenomenon of fluid turbulence in classical physics. A key parameter for studying turbulence is the Reynolds Number $UL/\nu$ where $U$ is a typical velocity,  $L$ is a length scale and $\nu$ molecular viscosity. By studying flows at large Reynolds Number, the dynamical equations have an almost precise balance between inertial and pressure gradient forces - irreversible viscous forces are negligible. Hence if $\nu$ is small, and the focus of attention is on laboratory scales, one can in practice neglect the role of viscosity in determining the motion of the fluid. Similarly, if state-space convergence of trajectories only occurs on the Planck scale, then on ordinary laboratory scales, one can safely neglect its effect in computing quantum physical effects using Hamiltonian i.e. Unitary dynamics. 
 
The next phase of IST development is the construction of a (special) relativistic structure which will link directly to the Dirac equation. It has already been noted that the quaternionic structure of IST appears particularly compatible with the spinorial structure of the Dirac equation. From there one can envisage describing the role of the gauge groups $U(1) \times SU(2) \times SU(3)$ on the symbolic sequences and from there perhaps construct a realistic form for the Standard Model. This is work for the future. 
 
 \section{Conclusions}
 \label{conclusions}
 
 A theory (IST) of multi-qubit systems has been developed which provides a basis for reinstating the type of physics in which Einstein believed passionately: a physics that is realistic deterministic and locally causal. Some physicists might be inclined to say: so what? Quantum theory has never been shown to be wanting experimentally, and there is therefore no compelling reason to want to change it. So in this concluding section, three possible areas where IST may show some practical advantage over quantum theory are outlined. 
 
The first is in the area of entanglement. Quantum entanglement is at the heart of many tasks in quantum information and quantum computing \cite{Jozsa:1997}. However, apart from simple cases (low dimensions, few particles, pure states), however, the mathematical structure of entanglement is not yet fully understood and is notoriously hard to characterise. That barriers to understanding exist can be appreciated if one considers the use of Hopf fibrations of the sphere as a geometric means to characterise the entangled states of 1, 2 and 3 qubits \cite{Mosseri:2001} \cite{BernevigChen:2004}. For example, the (normalised) Hilbert Space of a 1 qubit state is $\mathbb{S}^3$ which using the first Hopf fibration, can be written locally as $\mathbb{S}^2 \times \mathbb{S}^1$. Similarly, the Hilbert Space of a 2 qubit state is $\mathbb{S}^7$ which, from the second Hopf fibration, can be written locally as $\mathbb{S}^4 \times \mathbb{S}^3$. Finally, the Hilbert Space of a 3-qubit state is $\mathbb{S}^{15}$ which, from the third Hopf fibration, can be written locally as $\mathbb{S}^8 \times \mathbb{S}^7$. These Hopf fibrations are linked to the existence of algebraic structures: the complex numbers, quaternions and octonions respectively. However, there are no Hopf fibrations for $\mathbb{S}^n$, $n>15$, related to the fact that algebras higher than those of the octonions have zero divisors and therefore do not form division algebras. This prevents a similar geometric decomposition for 4 or more qubits. What is the physical implication of this?

From the perspective of IST, this may signal a potential problem for quantum theory, arising directly from the continuum (vector space) nature of Hilbert Space. As discussed in the section below, IST sees the continuum nature of the Hilbert Space as a computational convenience and not a reflection of physical reality. However, in IST there is a natural granular decomposition of lbit states associated with 4 or more qubits and entanglement can be simply represented by identifying the subscripts of the square-root-of-minus-one operators (see Section \ref{EI}). Hence, a possible advantage of IST over quantum theory is that the explicit lbit representations above may provide a more constructive characterisation of entanglement in quantum physics than is possible in quantum theory. 

Entanglement is the key quantum theoretic resource which allows certain quantum algorithms to be executed exponentially faster than their classical counterparts \cite{Jozsa:1997}. Although quantum computation has not been discussed in this paper, it can be seen that the invariant set $\mathcal{I}_D$ is a resource not available to a local (in state space) integration of $D$ itself. This resource is not available in classical dynamical systems whose equations of motion are based entirely on local differential or difference equations. Formally, fractal invariant sets are not computably related to $D$ \cite{Blum} and therefore equations which define the geometry of such invariant sets contains information that would need an infinitely long integration of $D$ to determine. Hence, if IST is indeed a viable theory of multi-qubit physics, it provides a fundamentally new insight into the origin of the exponential speed up of certain quantum computations over their classical counterparts, and it may be able to exploit this insight to advance both the theory and practice of quantum computation (e.g. in determining new classes of algorithm which have quantum efficiency). It might be interesting to analyse this possibility by treating chaotic dynamical systems as computational devices, with and without the constraint that the states of such systems lie on their fractal invariant sets. 

The second area where IST may show some advantage over quantum theory is in exploring the phenomenon of weak measurement. Measurement is such an intrinsic component of the foundations of quantum theory that the form of the Uncertainty Principle in quantum theory suggests that there is a linkage on the one hand between some intrinsic indeterminacy that a quantum system must possess, and on the other hand the impact that measurements have in disturbing a quantum system. Recent experimental results (\cite{Rozema:2012}) show that there is no such inherent linkage, and demonstrate a degree of measurement precision that can be achieved with weak-measurement techniques. Here it is suggested that a more penetrating analysis of weak measurement experiments will be possible in IST than can be achieved in standard quantum theory. 

Finally, IST predicts that there is no such thing as a `graviton'; if an experiment could ever be devised to detect gravitons, IST predicts such an experiment will give a null result. The key reason is that the basic physics which underpins IST is gravitational in nature. According to IST, the notion of the graviton as a quantum excitation of the gravitational field, obtained by applying quantum field theory to some classical gravitational theory, is a misguided one. 

\bibliography{mybibliography}

\begin{thebibliography}{10}

\bibitem{Alexander:1992}
J.~Alexander, J.A.Yorke, Z.~You, and I.Kan.
\newblock Riddled basins.
\newblock {\em Int. J. Bif. Chaos}, 2:795, 1992.

\bibitem{GrabenAtm}
P.~beim Graben and H.~Atmanspacher.
\newblock Complementarity in classical dynamical systems.
\newblock {\em Foundations of Physics}, 36:291--306, 2006.

\bibitem{Bell}
J.S. Bell.
\newblock {\em Speakable and unspeakable in quantum mechanics}.
\newblock Cambridge University Press, 1993.

\bibitem{BernevigChen:2004}
B.~Bernevig and H.-D.Chen.
\newblock Geometry of the 3-qubit state, entanglement and division algebras.
\newblock {\em arXiv:quant-ph/0302081}, 2004.

\bibitem{Blum}
L.~Blum, F.Cucker, M.Shub, and S.Smale.
\newblock {\em Complexity and Real Computation}.
\newblock Springer, 1997.

\bibitem{BohmHiley}
D.~Bohm and B.J.Hiley.
\newblock {\em The Undivided Universe}.
\newblock Routledge, 2003.

\bibitem{DoeringGibbon}
C.R.Doering and J.D.Gibbon.
\newblock {\em Applied Analysis of the Navier-Stokes Equation}.
\newblock Cambridge University Press, 1995.

\bibitem{Diosi:1989}
L.~Di\'{o}si.
\newblock Models for universal reduction of macroscopic quantum fluctuations.
\newblock {\em Phys. Rev.}, A40:1165--74, 1989.

\bibitem{Dube:1993}
S.~Dube.
\newblock Undecidable problems in fractal geometry.
\newblock {\em Complex Systems}, 7:423--444, 1993.

\bibitem{Gilmore}
R.~Gilmore and M.~Lefranc.
\newblock {\em The Topology of Chaos}.
\newblock Wiley, 2002.

\bibitem{Jahnel:2005}
J.~Jahnel.
\newblock {\em When does the (co)-sine of a rational angle give a rational
  number?}
\newblock www.uni-math.gwgd.de/jahnel/linkstopapers.html, 2005.

\bibitem{Maldacena:2004}
J.M.Maldacena.
\newblock {\em Quantum gravity as an ordinary gauge theory. In Science and
  Ultimate Reality: Quantum Theory, Cosmology and Complexity. Eds J.D Barrow,
  P.C.W. Davies and C.L. Harper Jr.}
\newblock Cambridge University Press, 2004.

\bibitem{Jozsa:1997}
R.~Jozsa.
\newblock {\em Entanglement and Quantum Computation. Appearing in `Geometric
  Issues in the Foundations of Science' eds. S. Huggett, L. Mason, K. P. Tod,
  S. T. Tsou and N. M. J. Woodhouse}.
\newblock Oxford University Press, 1997.

\bibitem{LindMarcus}
D.~Lind and B.~Marcus.
\newblock {\em An Introduction to Symbolic Dynamics and Coding}.
\newblock Cambridge University Press, 1995.

\bibitem{Lloyd:2012}
S.~Lloyd.
\newblock A turing test for free will.
\newblock {\em Phil. Trans. Roy. Soc.}, A370:3597--3610, 2012.

\bibitem{Palmer:1978}
T.N. Palmer.
\newblock Covariant conservation equations and their relation to the energy
  momentum concept in general relativity.
\newblock {\em Phys.Rev.}, D18:4399 4407, 1978.

\bibitem{Palmer:1980}
T.N. Palmer.
\newblock Gravitational energy momentum: The einstein pseudo-tensor re
  examined.
\newblock {\em Gen.Rel. and Gravitation}, 12:149--154, 1980.

\bibitem{Palmer:2009a}
T.N. Palmer.
\newblock The invariant set postulate: a new geometric framework for the
  foundations of quantum theory and the role played by gravity.
\newblock {\em Proc. Roy. Soc.}, A465:3165--3185, 2009.

\bibitem{Penrose:2004}
R.~Penrose.
\newblock {\em The Road to Reality: A Complete Guide to the Laws of the
  Universe}.
\newblock Jonathan Cape, London, 2004.

\bibitem{Penrose:2010}
R.~Penrose.
\newblock {\em Cycles of Time: An Extraordinary New View of the Universe}.
\newblock The Bodley Head, 2010.

\bibitem{Penrose:2011}
R.~Penrose.
\newblock Uncertainty in quantum mechanics: faith or fantasy.
\newblock {\em Phil. Trans. Roy. Soc.}, A369:4679--4937, 2011.

\bibitem{Mosseri:2001}
Mosseri R. and R.~Dandoloff.
\newblock Geometry of entangled states, bloch spheres and hopf fibrations.
\newblock {\em J. Phys. A: Math. Gen.}, 34:10243--10252, 2001.

\bibitem{Rovelli}
C.~Rovelli.
\newblock {\em Quantum Gravity}.
\newblock Cambridge University Press, 2004.

\bibitem{Rozema:2012}
L.~A. Rozema, A.~Darabi, D.H. Mahler, A.~Hayat, Y.~Soudagar, and A.~M.
  Steinberg.
\newblock Violation of heisenbergÕs measurement-disturbance relationship by
  weak measurements.
\newblock {\em Phys. Rev. Lett.}, 109:100404--9, 2012.

\bibitem{Schwinger}
J.~Schwinger.
\newblock {\em Quantum Mechanics: Symbolism of Atomic Measurements}.
\newblock Springer, 2001.

\bibitem{Wainwright}
J.~Wainwright and G.F.R. Ellis.
\newblock {\em Dynamical System in Cosmology}.
\newblock Cambridge University Press, 1997.

\bibitem{Williams}
S.G. Williams.
\newblock {\em Symbolic Dynamics and its Applications}.
\newblock American Mathematical Society, 2004.

\bibitem{tHooft}
G~Õt~Hooft.
\newblock {\em Class.Quant.Grav.}, 16:3263--3279, 1999 see also gr-qc/9903084.

\end{thebibliography}

\appendix
\section{When does the cosine of a rational angle give a rational number?}

$\mathbf{Theorem}$\cite{Jahnel:2005}.  Let $0 < \theta/\pi < 1/2 \in \mathbb{Q}_2$. Then $\cos \theta \notin \mathbb{Q}$. 

We derive a \emph{reductio ad absurdum}. Assume that $\cos \theta = a/b$ is rational, where $a, b \in \mathbb{Z}, b \ne 0$ have no common factors.  Using the identity $2 \cos 2\theta = (2 \cos \theta)^2-2$ we have
\be
2\cos 2\theta = \frac{a^2-2b^2}{b^2}
\ee
Now $a^2-2b^2$ and $b^2$ have no common factors, since if $p$ were a prime number dividing both, then $p|b^2 \implies p|b$ and $p|(a^2-2b^2) \implies p|a$, a contradiction. Hence if $b \ne \pm1$, then the denominators in $2 \cos \theta, 2 \cos 2\theta, 2 \cos 4\theta, 2 \cos 8\theta \dots$ get bigger and bigger without limit. On the other hand, with $0 < \theta/\pi < 1/2 \in \mathbb{Q}$, then $\theta/\pi=m/n$ where $m, n \in \mathbb{Z}$ have no common factors. This implies that the sequence $(2\cos 2^k \theta)_{k \in \mathbb{N}}$ admits at most $n$ values. Hence we have a contradiction. Hence $b=\pm 1$ and $\cos \theta =0, \pm1/2, \pm1$. No $0 < \theta/\pi < 1/2 \in \mathbb{Q}_2$ has $\cos \theta$ with these values. 
 
 \end{document}